\documentclass[times]{weauth}
\usepackage{moreverb}

\usepackage{subcaption}
\usepackage{multirow}
\usepackage{cite}

\newcommand\BibTeX{{\rmfamily B\kern-.05em \textsc{i\kern-.025em b}\kern-.08em
T\kern-.1667em\lower.7ex\hbox{E}\kern-.125emX}}

\newcommand{\bs}{\boldsymbol}

\def\volumeyear{2017}

\begin{document}

\runningheads{L.~J.~Lukassen et al.}{Space-time correlations in wind farms}

\articletype{\noindent This manuscript has been submitted to Wind Energy. \nobreak}

\title{Modeling space-time correlations of velocity fluctuations in wind farms}

\author{Laura J.~Lukassen\affil{1}, Richard J.~A.~M.~Stevens\affil{2}, Charles Meneveau\affil{3} and Michael Wilczek$^1$}

\address{\affilnum{1} Max Planck Institute for Dynamics and Self-Organization, Goettingen, Germany\\
\affilnum{2} Department of Science and Technology and J.M. Burgers Center for Fluid Dynamics, University of Twente, The Netherlands\\
\affilnum{3} Department of Mechanical Engineering, Johns Hopkins University, Baltimore, USA}

\corraddr{Michael Wilczek, Max Planck Institute for Dynamics and Self-Organization, Am Fassberg 17, 37077 Goettingen, Germany.\\ Email: michael.wilczek@ds.mpg.de}

\begin{abstract}
An analytical model for the streamwise velocity space-time correlations in turbulent flows is derived and applied to the special case of velocity fluctuations in large wind farms. The model is based on the Kraichnan-Tennekes random sweeping hypothesis, capturing the decorrelation in time while including a mean wind velocity in the streamwise direction. In the resulting model, the streamwise velocity space-time correlation is expressed as a convolution of the pure space correlation with an analytical temporal decorrelation kernel. Hence, the spatio-temporal structure of velocity fluctuations in wind farms can be derived from the spatial correlations only. We then explore the applicability of the model to predict spatio-temporal correlations in turbulent flows in wind farms. Comparisons of the model with data from a large eddy simulation of flow in a large, spatially periodic wind farm are performed, where needed model parameters such as spatial and temporal integral scales and spatial correlations are determined from the large eddy simulation. Good agreement is obtained between the model and large eddy simulation data showing that spatial data may be used to model the full temporal structure of fluctuations in wind farms.
\end{abstract}


\maketitle

\section{Introduction}
\label{sec:introduction}
The wind energy market is strongly growing with a record in new installations in 2015 \cite{WWEAMarch2016, WWEAonline} and increasing importance expected over the next years, see also \cite{WindEnergyScenarios2030}. Wind energy is a fluctuating resource that is converted into electrical power in a nonlinear fashion (see e.g. \cite{Burton2011}). These characteristics constitute challenges with respect to power quality and grid stability, cf.~\cite{MilanPeinkePRL2013}, which raise significant interest in predicting short-term power output fluctuations in wind farms. Understanding and predicting the spatio-temporal structure of velocity fluctuations in wind farms which is a main source of the power output fluctuations, is a crucial step towards this goal.

The atmospheric conditions of wind approaching a wind farm can be described by basic atmospheric boundary layer (ABL) theory. For a recent overview, we refer to \cite{StevensMeneveau2017Annu}. 
Wind speed fluctuations in the ABL occur on different time scales, ranging from annual or seasonal variations, down to minutes and seconds \cite{Burton2011, vanderHoven}. 
The short-term fluctuations on scales smaller than around, say, ten minutes are usually referred to as turbulent fluctuations \cite{Burton2011}.
An introduction to the characteristics of atmospheric wind including typical turbulence intensities and naturally occurring extreme wind gusts (cf.~also \cite{BoettcherBarthPeinke2007}), as well as other factors such as geographical (onshore, offshore), landscape and climate influences is given in \cite{Burton2011}.
The evolution of wind velocities in the wind farm itself is influenced by the aerodynamics of the turbines (see e.g.~\cite{HansenButterfieldAnnRevFM1993}), the design of the whole wind farm, e.g.~in terms of turbine spacing \cite{MeyersMeneveauWindEnergy2012, StevensGaymeMeneveauWindEnergy2016}, and by the influence of turbines onto each other, primarily mediated by turbine wakes \cite{CrespoHernandezFrandsenWindEnergy1999, VermeerSorensenCrespo2003, KeaneAguirreorque2016, FletcherBrownWindEnergy2010, SamoraniHandbook}. Capturing the full complexity of turbulent flows in wind farms at all scales simultaneously is  currently computationally out of reach, emphasizing the need for simplified, physics-based models.

In the present work, we derive such a model for the space-time correlation of the turbulent short-term fluctuations based on concepts from classical turbulence theory. Specifically, it is based on the Kraichnan-Tennekes random sweeping hypothesis with additional mean flow. Focusing on its simplicity, it depends on a minimal set of parameters and is set up to be applied both in the ABL without turbines, or in the wind turbine array boundary layer. With respect to wind energy applications, it provides a starting point for further developments, such as predictive models of the temporal power output fluctuations, or in the area of wind turbine control \cite{Bossanyi2013}.

Velocity space-time correlations along with modeling approaches have been widely investigated (see \cite{Favre1965} for a classical and \cite{HeJinYangAnnuRev2017,Wallace2014} for recent reviews) and have also found application to turbulent ABLs. Thus, applying similar approaches to the modeling of wind farms appears worth exploring, which is the scope of the present paper.

Considering turbulence with mean flow, velocity fluctuations are primarily advected with the mean velocity, which forms the basis of Taylor's frozen eddy hypothesis \cite{Taylor}. Due to the frozen-in assumption, however, these fluctuations do not decorrelate in time along the mean flow. In the random sweeping hypothesis, as introduced by Kraichnan \cite{Kraichnan} and Tennekes \cite{Tennekes}, it is assumed that small-scale velocity perturbations are advected randomly by large-scale velocity fluctuations. This random advection is the primary source of temporal decorrelation. Both Taylor's and the Kraichnan-Tennekes hypotheses have been thoroughly investigated with respect to their range of validity, e.g.~in \cite{Lin1953,Lumley1965, WuGengYao2017, KatulBanerjee2016}. Due to their simplicity, they have laid the foundation for a range of models for space-time correlations both in real space \cite{HeZhang2006,ZhaoHe2009} as well as in spectral space \cite{HuntBuellWray1987,WilczekNarita2012}. In the spectral formulation of space-time correlations in terms of wavenumber-frequency spectra, the random sweeping hypothesis can be applied to derive the wavenumber-frequency spectrum as a product of the wavenumber spectrum and a frequency distribution, whose properties are fixed by the decorrelation hypothesis \cite{WilczekNarita2012}. In a real space formulation, as adopted in this paper, this corresponds to a convolution of the spatial correlation function with a temporal decorrelation kernel, consistent with the classical  Kovasznay-Corrsin conjecture \cite{Corrsin, Favre1965, Phillips2000}. Recently, the random sweeping model with additional mean flow advection has been extended to include the physics of ABLs to establish a spectral description of ABL space-time correlations \cite{WilczekStevensMeneveau2015}.

In the context of ABLs and wind farms, a variant of space-time correlations, so-called coherence functions, are widely used which contain space-frequency information. Modeling approaches usually combine Taylor's frozen eddy hypothesis with a model for the spectrum of wind speed fluctuations, such as the Kaimal spectrum \cite{Kaimal1972} or the von K\'arm\'an spectrum \cite{vonKarman1948}, see \cite{Burton2011}. As an example, in the Mann model of turbulence \cite{Mann1994, Mann1998}, two-point spatial information is obtained on the basis of one-point spectra with the help of Taylor's frozen eddy hypothesis. Instead of using Taylor's frozen eddy hypothesis, de Mar\'{e} and Mann \cite{MareMann2016} combine the Mann model mentioned above with a model by Kristensen \cite{Kristensen1979} to model space-time correlations in high-Reynolds number flow. Kristensen's longitudinal coherence is determined by the temporal decay of eddies and their transverse motion. Other coherence models which also avoid the use of Taylor's frozen eddy hypothesis can be found in literature, e.g.~\cite{SoerensenHansenRosas2002, Bossanyi2013}.

In the interest of a broader overview, we also mention stochastic modeling approaches which are used to predict velocity increment statistics in wind farms \cite{MeliusTutkunCal2014n1, MeliusTutkunCal2014n2,MilanPeinkePRL2013}, and wind power output fluctuations, e.g.~\cite{GottschallPeinke2007, GottschallPeinke2008, AnahuaBarthPeinke2008}. A description of wind power output fluctuations influenced by atmospheric turbulence can be found in \cite{Bandi2017}. 

The model for space-time correlations in wind farms based on the random sweeping hypothesis derived in this paper essentially represents a real space formulation of the previous spectral model for ABL introduced in \cite{WilczekStevensMeneveau2015}, generalized here to include finite time correlations of the random sweeping velocity. Given that the flow through a wind farm is strongly inhomogeneous due to the presence of the wind turbines, a real-space formulation appears favorable as it allows us to discern different flow regions such as along and between streamwise lines of turbines. The finite-time correlation of the random advection velocity is crucial for capturing the decorrelation trends at large times more accurately.

The remainder of this paper is structured as follows. In section \ref{sec:Theory}, the mathematical derivation of the model in real space is presented. In a next step, our model is compared to data obtained from large eddy simulation (LES) of turbulent flow in a wind farm. In section \ref{sec:LES}, we briefly describe the LES tool and simulated flow conditions. The comparison of the model to the numerical data is given in section \ref{sec:Comparison}. The principal aim of the analysis and comparison is to establish whether the space-time correlation model in terms of a convolution of the space correlation and a temporal decorrelation kernel can predict observations made from the LES wind farm data. Further details are discussed in an outlook, section \ref{sec:Outlook}.

\section{Space-time correlations from an advection model}
\label{sec:Theory}
In the following, we provide a derivation for the space-time correlations of streamwise velocity fluctuations $u_1'$ in a plane at hub height in wind farms. It is based on the Kraichnan-Tennekes random sweeping hypothesis \cite{Kraichnan, Tennekes}; we assume that small-scale velocity fluctuations are advected with a large-scale random velocity $\bs v=(v_1, v_2)$, which we, for simplicity, restrict to the horizontal plane. The indices $1$ and $2$ refer to the streamwise and spanwise direction, respectively. Additionally, we take into account an advection with a mean velocity $\bs U = (U,0)$ in streamwise direction, corresponding to the mean wind in an ABL. For now, we neglect possible differences between the mean velocity and effective convection velocity, a topic of considerable interest in its own right, see e.g.~\cite{Wills1964,kimhussain1993, delalamojimenez2009} for turbulent shear flows. The simple dynamical model discussed in the following serves to motivate the analytical structure of the space-time correlations. We then conjecture that, in essence, the same model holds for the arguably more complex flow in the wind farm and compare the results to LES.

\subsection{Advection model}
Assuming that advection with $\bs U$ and $\bs v$ gives the dominant contribution to capture the temporal decorrelation of velocity fluctuations, we arrive at the simple advection equation for the streamwise small-scale velocity component:
\begin{equation}\label{eq:advection}
  \partial_t u_1' + (\bs U+ \bs v) \cdot \nabla u_1' = 0 \, .
\end{equation}
To include the random character of the large-scale advection, we assume that $\bs v$ is spatially constant with a Gaussian ensemble distribution and decorrelates exponentially in time:
\begin{equation}
 \langle v_1(t)v_1(t+\tau) \rangle = \langle v_1^2 \rangle \exp\left[-   |\tau| /T_1 \right]\, ,
 \label{eq:exp_decay_rsvel}
\end{equation}
for the streamwise random-sweeping velocity $v_1$, and analogously for the spanwise random-sweeping velocity $v_2$. Here, $\langle \dots \rangle$ denotes ensemble-averaging. For practical purposes in the application to wind farm LES data, the ensemble averages are replaced by averages over space and time which is further discussed in the corresponding sections. The assumption of an exponentially decaying correlation function allows us to introduce decorrelation time scales $T_1$ and $T_2$, respectively. This assumption is tested below with our LES data. 
In the following, we explicitly compute the space-time correlations of the streamwise velocity
\begin{equation}
 R_{11}(\bs r,\tau)  =  \left\langle u_1'(\bs x+\bs r,t+\tau) \, u_1'(\bs x,t) \right\rangle
 \label{eq:R_equation1}
\end{equation}
in the framework of the advection model, where $\bs r = (r_1, r_2)$ is the distance vector in the plane at hub height and $\tau$ is the time lag. The solution of equation \eqref{eq:advection} corresponds to a shifted initial condition such that
\begin{equation}
  u_1'(\bs x+\bs r,t+\tau) = u_1'\left(\bs x+\bs r - \bs U\tau - \int_t^{t+ \tau } \!\!\!\!\! \bs v(s) \, \mathrm{d}s,t \right) \, . 
  \label{eq:u_shifted}
\end{equation}
Under the assumption of homogeneity and statistical stationarity, there is no explicit $\bs x$- and $t$-dependence on the left hand side of equation \eqref{eq:R_equation1}. Using equation \eqref{eq:u_shifted}, the space-time correlations can be obtained according to
\begin{align}
 R_{11}(\bs r,\tau)  &= \left\langle u_1'\left(\bs x + \bs r - \bs U\tau - \int_t^{t+ \tau} \!\!\!\!\! \bs v(s) \, \mathrm{d}s,t \right) \, u_1'(\bs x,t) \right\rangle \label{eq:model_space_time_corr1} \\
		&=  \int\mathrm{d}\bs a \left\langle u_1'\left(\bs x+\bs r-\bs a,t \right) \, u_1'(\bs x,t) \, \delta\left(\bs a-\bs U\tau- \int_t^{t+ \tau } \!\!\!\!\! \bs v(s) \, \mathrm{d}s \right) \right\rangle \label{eq:model_space_time_corr2} \\
		&=  \int\mathrm{d}\bs a \left\langle u_1'\left(\bs x+\bs r-\bs a,t \right) \, u_1'(\bs x,t) \right\rangle\left\langle \delta\left(\bs a-\bs U\tau- \int_t^{t+ \tau } \!\!\!\!\! \bs v(s) \, \mathrm{d}s \right) \right\rangle \, . \label{eq:model_space_time_corr3}
\end{align}
The average here is taken over both, small-scale and large-scale fluctuations. To factorize the averages in the last step, we have assumed statistical independence of the large-scale random sweeping velocity $\bs v$ at all times in the interval $[t, t+\tau]$ and the streamwise velocity fluctuations $u_1'$. A similar assumption has been made by Corrsin in the context of the relation between Eulerian and Lagrangian velocity correlations \cite{Corrsin , Phillips2000}. As a result of this factorization, the first average in the integrand is simply the instantaneous correlation function evaluated at $\bs r-\bs a$ in the plane, i.e.~$R_{11}(\bs r-\bs a)$. The averaged delta function represents the decorrelation kernel which determines the temporal decorrelation of the velocity fluctuations. Under the Gaussian assumption for the large-scale velocity, it can be explicitly evaluated yielding
\begin{equation}
  \left\langle \delta\left(\bs a-\bs U\tau- \int_t^{t+ \tau } \!\!\!\!\! \bs v(s) \, \mathrm{d}s \right) \right\rangle = \frac{1}{2\pi \sigma_1(\tau)\sigma_2(\tau)}\exp\left[ -\frac{(a_1-U\tau)^2}{2\sigma_1(\tau)^2} -\frac{a_2^2}{2\sigma_2(\tau)^2} \right]\, ,
  \label{eq:decorrelation_kernel}
\end{equation}
where
\begin{align}
  \sigma_1(\tau)^2 &= \int_t^{t+\tau}\!\!\!\mathrm{d}s \int_t^{t+ \tau }\!\!\!\mathrm{d}s'  \left\langle v_1(s) v_1(s') \right\rangle = 2\langle v_1^2 \rangle T_1^2 \left[ \frac{ |\tau| }{T_1} - \left( 1-\exp\left[ -\frac{ |\tau | }{T_1} \right] \right) \right]\, , \label{eq:sigma_1} \\
    \sigma_2(\tau)^2 &= \int_t^{t+ \tau }\!\!\!\mathrm{d}s \int_t^{t+ \tau }\!\!\!\mathrm{d}s'  \left\langle v_2(s) v_2(s') \right\rangle = 2\langle v_2^2 \rangle T_2^2 \left[ \frac{ | \tau | }{T_2} - \left( 1-\exp\left[ -\frac{ | \tau |}{T_2} \right] \right) \right] \, . \label{eq:sigma_2}
\end{align}
The effect of the Gaussian decorrelation kernel on the space-time correlations can be interpreted straightforwardly: the mean velocity advection induces a shift of the spatial correlation function, whereas the random advection blurs the spatial correlation by mixing a range of spatial scales. The functional forms of $\sigma_1$ and $\sigma_2$ are implied by the exponential decay of the large-scale velocities. While these formulas are already quite involved, the essential point here is that $\sigma_1(\tau)^2 \approx \langle v_1^2 \rangle \tau^2$ for $ |\tau |  \ll T_1$ and $\sigma_1(\tau)^2 \approx 2\langle v_1^2 \rangle T_1  |\tau |$ for $| \tau |   \gg T_1$, i.e.~the random displacement exhibits a ballistic behavior for short times and then transitions to a diffusive long-time behavior. This behavior, in fact, is not limited to exponentially decaying correlation functions but can be realized with a much larger class of correlation functions. 

In the limit of $|\tau| \rightarrow 0$, the decorrelation kernel becomes a delta function; as expected, the space-time correlation function reduces to the spatial correlation only, $R_{11}(\bs r,0) = R_{11}(\bs r)$. For large time increments, $|\tau| \rightarrow \infty$, the decorrelation kernel becomes very broad and shallow, which leads to a vanishing correlation in time.
 
\subsection{Reduction to a one-dimensional model}
\label{subsec:Reduction}
Because we aim for an even simpler, one-dimensional model for the streamwise velocity correlation in the streamwise direction $r_1$, we make the assumption that the spatial correlation function in the plane can be factorized according to $R_{11}(\bs r) = R_{11}(r_1) \exp\left[-r_2^2/\left(\frac{4}{\pi} L_2^2\right) \right]$, i.e.~the transverse decorrelation of the streamwise velocity is modeled as a Gaussian with a transverse integral length scale $L_2$. Throughout the paper, we use the convention that functions such as $R_{11}(...)$ only contain those parameters as arguments which are not a priori set to zero. The factoring assumption, which is robust with respect to the functional form of the transverse decorrelation, allows to explicitly evaluate the transverse contribution of the convolution integral. As a result, we obtain
\begin{equation}\label{eq:spacetimecorr}
R_{11}(r_1,\tau) = \int\!\mathrm{d}a_1 \, R_{11}(r_1-a_1) \frac{\exp\left[ -\frac{(a_1-U\tau)^2}{2\sigma_1(\tau)^2} \right]}{\sqrt{2\pi \sigma_1(\tau)^2 \left[ 1 + \sigma_2(\tau)^2/\left(\frac{2}{\pi} L_2^2\right)\right]}}  \, .
\end{equation}
The space-time correlation is given by a convolution of the instantaneous correlation function with a Gaussian temporal decorrelation kernel, which is parameterized by the mean velocity and the mean square random sweeping velocities. Interestingly, the width of the decorrelation kernel is given by the effect of streamwise random sweeping $\sigma_1$, whereas its amplitude depends on both, streamwise and spanwise random sweeping. Furthermore, $\sigma_2(\tau)$ and $L_2$ only appear in combination, $\sigma_2(\tau)^2/\left(\frac{2}{\pi} L_2^2\right)$, and not independently of each other. 

The derivation of our model resembles parts of the derivation of the wavenumber-frequency model for ABLs \cite{WilczekStevensMeneveau2015}, however, it is further extended by incorporating the temporal decorrelation of the random sweeping. Our resulting model also shows similarities to the Kovasznay-Corrsin convolution as presented in \cite{Phillips2000} for homogeneous turbulence advecting in streamwise direction. 

To sum up, the model depends upon the following parameters: the advection velocity $U$, the second moment of the streamwise and spanwise random sweeping velocities $\langle v_1^2 \rangle$, $\langle v_2^2 \rangle$, the correlation time scales of the random sweeping velocities $T_1$, $T_2$, the transverse integral length scale of the streamwise velocity fluctuations $L_2$ and the space correlation of the streamwise velocity fluctuations $R_{11}(r_1)$. By a comparison to LES, we explore the potential of the model presented here to describe space-time correlations of the streamwise velocity fluctuations in the hub height plane in wind farms. In the following, we discuss how the free parameters can be determined from LES. As an alternative, these parameters could also be prescribed by further modeling assumptions, which is beyond the scope of the present paper.

\section{Large eddy simulations and parameters}
\label{sec:LES}
The applicability of the model in equation \eqref{eq:spacetimecorr} for describing space-time velocity correlations in wind farms is investigated by LES data.
The goal is to compare the velocity space-time correlations predicted by the model to velocity space-time correlation data from the LES.
\subsection{Description of LES}
In the wind farm LES, we solve the filtered incompressible Navier-Stokes equations without buoyancy, system rotation or other effects to model a neutral pressure-driven ABL flow over a rough surface \cite{CalafMeneveauMeyers2010,StevensMeneveauRenSus2014}. In horizontal directions, a pseudo-spectral discretization with periodic boundary conditions is adopted and second-order centered finite differences are used in the vertical direction. The subgrid stresses are modeled using a scale-dependent Lagrangian dynamic approach in conjunction with the Smagorinsky model and a sharp spectral cutoff test-filter \cite{bou05}. The wall stress at the ground is captured by a standard rough-wall model \cite{moe84} using velocities test-filtered at twice the grid scale \cite{bou05}. For the top boundary we use a zero-vertical-velocity and zero-shear-stress boundary condition. Therefore, the modeled flow effectively corresponds to a `half-channel flow' with an impermeable centerline boundary. Time-stepping is performed with a second-order Adams-Bashforth scheme. Here, we consider the idealized case of a fully developed wind turbine array boundary layer, modeled by a periodic array of wind turbines. The individual wind turbines are modeled as actuator disks  \cite{jim07,CalafMeneveauMeyers2010,ste14}. Their back-reaction on the flow is modeled by a thrust force added to the momentum equation. For a more complete account on wind farm modeling by means of LES, we refer the reader to \cite{meh14,StevensMeneveau2017Annu}.

Throughout the paper, $x_1$ and $x_2$ denote the streamwise and spanwise directions, respectively, and $x_3$ denotes the vertical direction. The simulation is performed on a $512 \times 256 \times 128$ grid discretizing a domain of size $L_{x_1}/H \times L_{x_2}/H \times L_{x_3}/H = 4\pi \times 2\pi \times 1$, where $H$ is the height of the ABL, i.e.~the domain height. In this domain, $16 \times 16$ turbines are placed with a turbine diameter $D = 0.1 \, H$ in an aligned configuration. The streamwise spacing of the turbines $S_{x_1}$ is $7.85\, D = 0.785\, H$, the spanwise spacing $S_{x_2}$ is $3.93 \, D =0.393 \, H$. The simulation is conducted with a time step of $6 \times 10^{-5}\, H/u^*$, where $u^*$ is the friction velocity associated with the driving pressure gradient. The data are collected after the statistically stationary state is reached. The velocity field at hub height ($x_3=0.1\, H$) is stored every ten time steps at $8192$ snapshots, thus, over a total duration of about $4.9 \, H/u^*$. This corresponds to about $3.5$ flow through times at hub height, or about $55$ inter-turbine travel times (i.e.~the time for traveling from one turbine to the next). 
\begin{figure}
\centering
  \includegraphics[trim =  1.0cm 0.0cm 0.5cm 0.0cm, clip, width=0.48\textwidth]
 {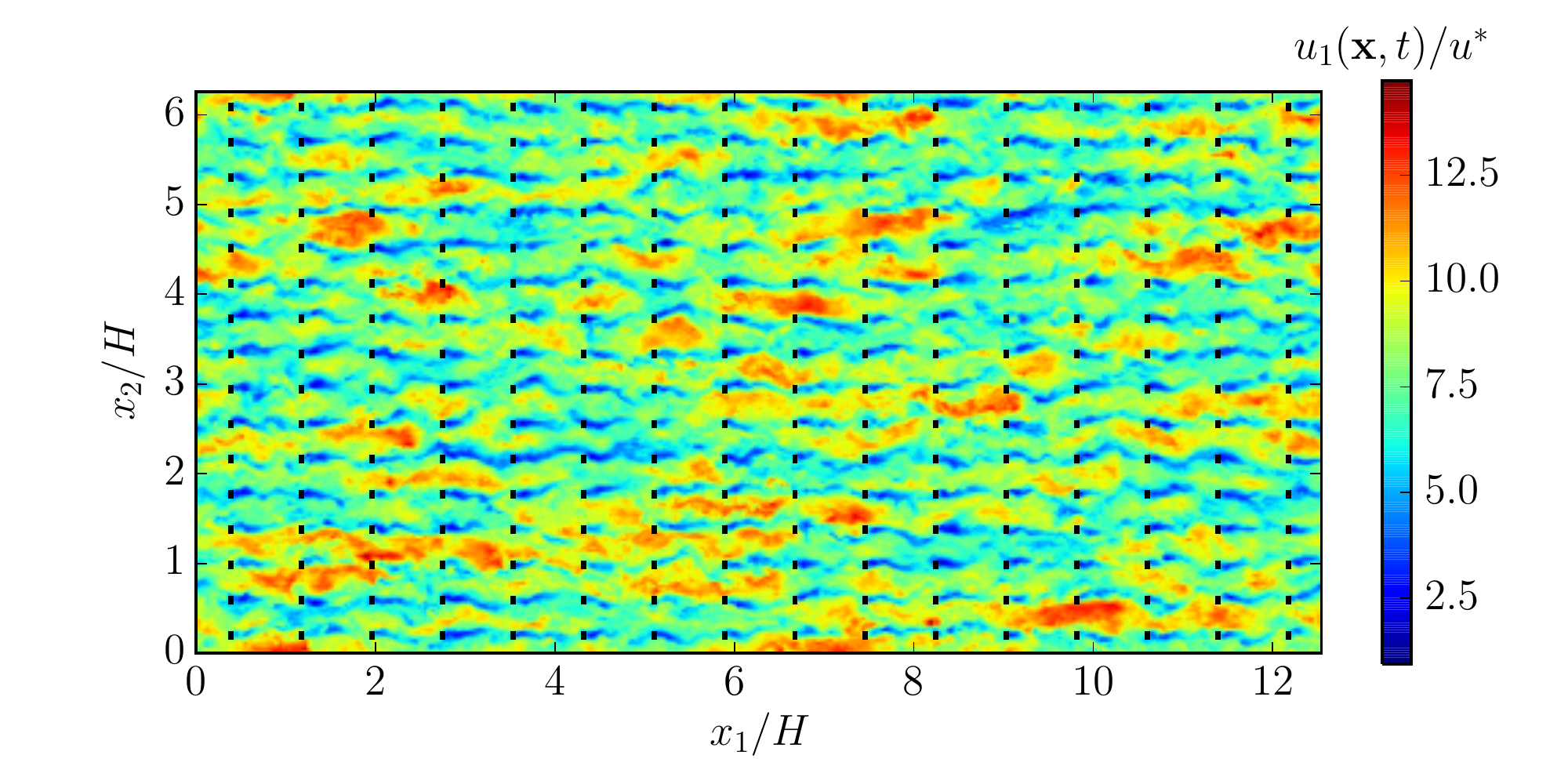}
\centering
  \includegraphics[trim =  1.0cm 0.0cm 0.5cm 0.0cm, clip, width=0.48\textwidth]
 {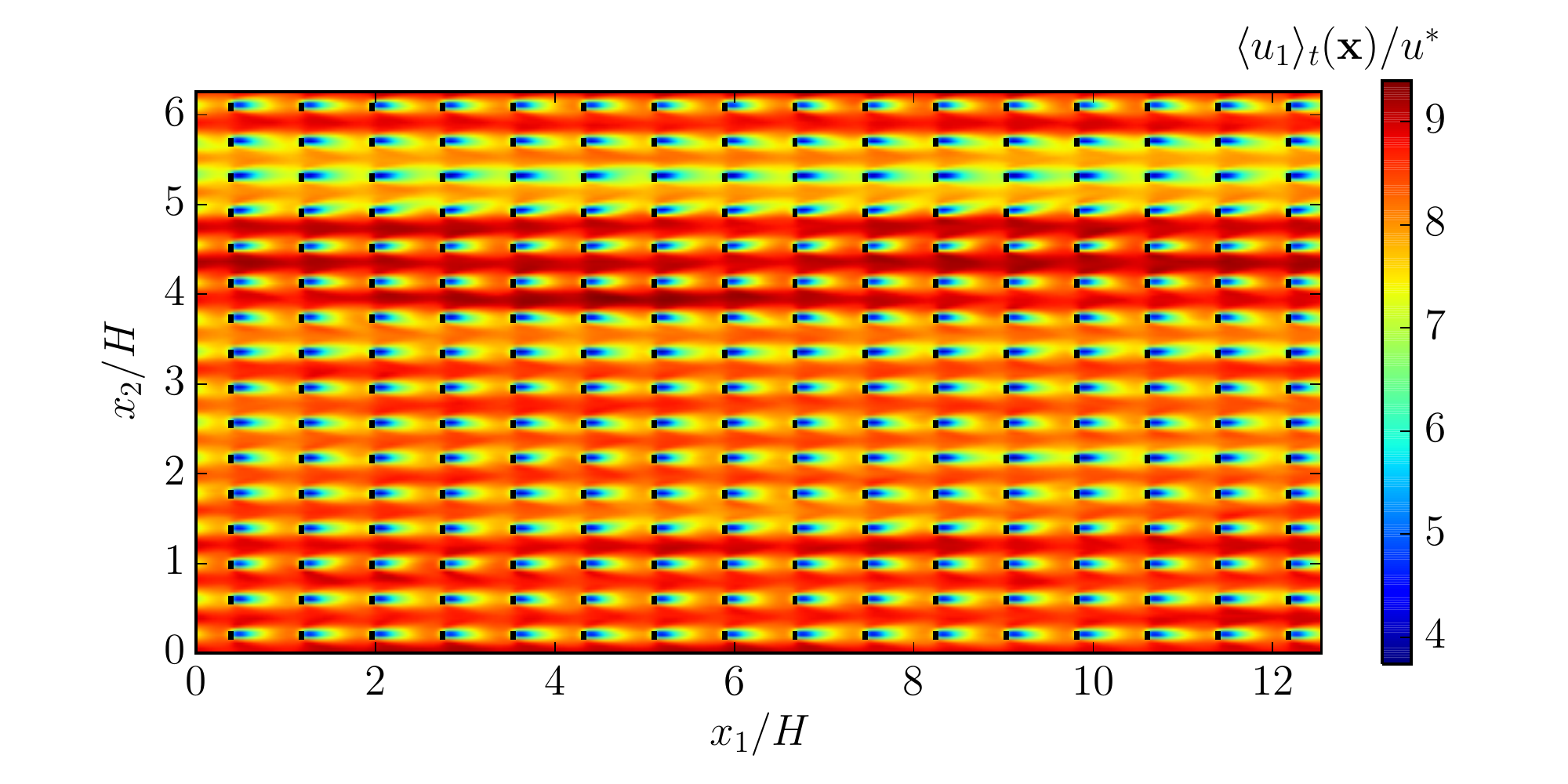} 
\caption{Left: Streamwise velocity component normalized by the friction velocity $u^*$ at one instant in time. $16 \times 16$ turbines are located in the domain indicated by the black lines. Right: Temporally averaged streamwise velocity component normalized by $u^*$. The individual turbines clearly introduce inhomogeneities in the flow field, especially along the lines of the turbines. One can also observe high-speed streaks between the turbines when averaging over time. All plots in this paper have been created with matplotlib \cite{Hunter2007}.
}
\label{fig:figure1_2}
\end{figure}
 \begin{figure}
  \centering
   \includegraphics[trim =1.0cm 0.0cm 0.5cm 0.0cm, clip, width=0.48\textwidth]
  {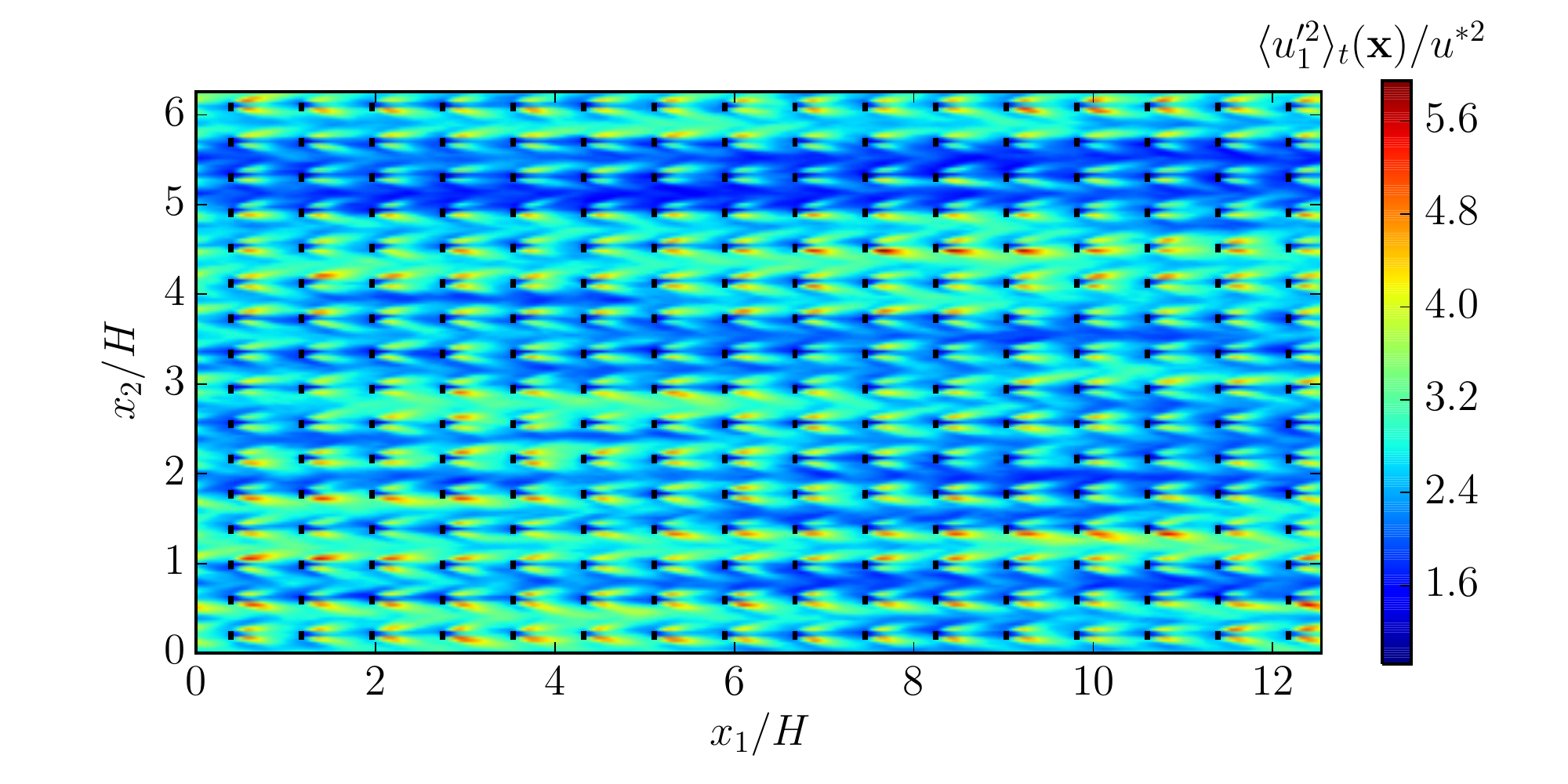}
\centering
\includegraphics[trim =  1.0cm 0.0cm 0.5cm 0.0cm, clip, width=0.48\textwidth] {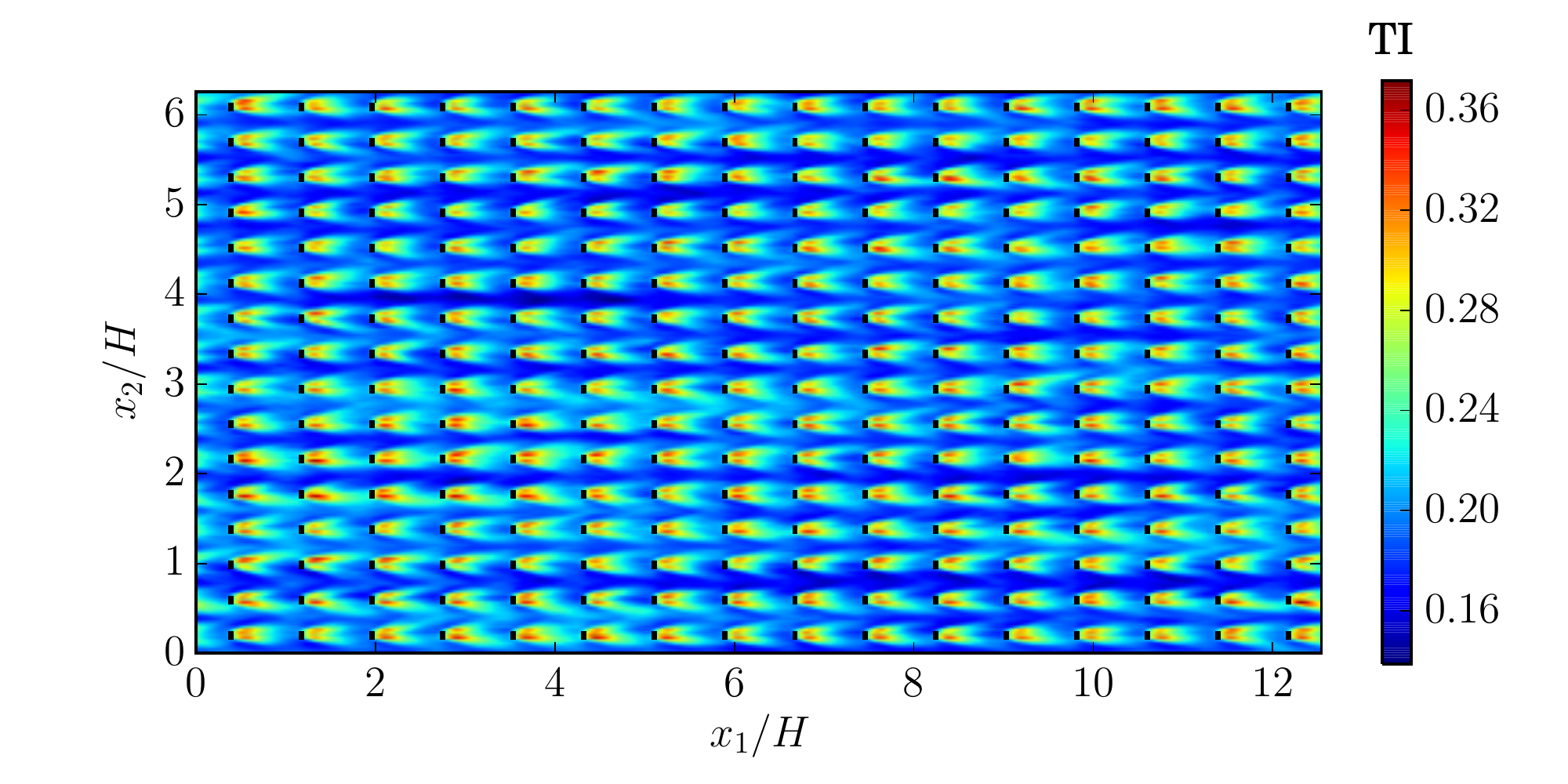}
\caption{Left: Temporally averaged second moment of the streamwise velocity fluctuations as given in equation \eqref{eq:time_average}, normalized by $u^{*2}$. The field is highly inhomogeneous. Right: Turbulence intensity of streamwise components, $\mathrm{TI} = \sqrt{\langle u_1'^2 \rangle_t (\bs x)}/\langle u_1 \rangle_t(\bs x)$. The highest intensity can be found at the edges of the turbine wakes.}
\label{fig:figure3_4}
\end{figure}
To give an impression of the flow, the left panel in figure \ref{fig:figure1_2} shows a visualization of the streamwise velocity. One can see that higher-velocity flow streaks meander between wind turbine columns. In the present paper, columns refer to turbines aligned in the streamwise direction, whereas rows refer to neighboring turbines in spanwise direction.
The right panel of figure \ref{fig:figure1_2} shows the temporal mean velocity. Even after time averaging there remain spatial large scale inhomogeneous streaks, visible as elongated regions. It is well known that such features are very slow to converge in this type of flow \cite{MuntersMeneveauMeyers2016}. Here, we average over these in space as appropriate, i.e.~when we investigate the applicability of the model between lines of turbines, we only average over all lines in the middle between the turbines in order to minimize the effect of these streaks. 
Furthermore, pronounced wakes behind the wind turbines are clearly visible.

For the subsequent analysis, we decompose the total streamwise velocity into its temporal mean and the remaining fluctuations
\begin{equation}
u_1(\bs x,t) = \langle u_1 \rangle_t(\bs x) + u_1'(\bs x,t)  \label{eq:decomposition} \, .
\end{equation}
The $\langle ... \rangle_t$-brackets denote an average over time. As a result, the temporal mean velocity is a function of the position in the plane. The mean square velocity fluctuations are obtained analogously:
\begin{equation}
\langle u_1'^2\rangle_t(\bs x) = \langle u_1^2 \rangle_t(\bs x)   -  \langle u_1 \rangle_t^2 (\bs x) \, .
\label{eq:time_average}
\end{equation}

The mean square velocity fluctuations are shown in the left panel of figure \ref{fig:figure3_4}.
As can be seen, the inhomogeneities introduced by the wind farm also carry over to the statistical quantities and have to be taken into account when computing statistical averages.
As expected, the temporal mean velocity is higher between the lines of wind turbines than along the lines, cf.~right panel in figure \ref{fig:figure1_2}. There, the velocity deficit behind the wind turbines is clearly visible.

As for the fluctuations, they appear lower between the turbine columns. The highest velocity fluctuations are generated at the edges of the wind turbine wakes. The turbulence intensity, defined as the relation of the root mean square fluctuations and the temporal mean velocity, is shown in the right panel of figure \ref{fig:figure3_4}. The highest turbulence intensity can be found at the edges of the turbine wakes accompanying the regions of the highest velocity fluctuations. Furthermore, it is visible that along the lines of the turbines, the highest intensity is located in the region behind the turbines within half the distance to the following turbines. This is followed by a less intense region in front of the following turbine.

\subsection{Determination of parameters and spatial correlations from LES}
\label{subsec:Determination from LES}

In the following, we compare the model space-time correlation to the space-time correlation obtained directly from the LES data. To this end, we compute the wavenumber-frequency spectrum from the LES data and determine the velocity space-time correlation by an inverse Fourier transform. To evaluate the model \eqref{eq:spacetimecorr}, we also compute the spatial correlation function. Based on the LES data, we determine values for the mean velocity $U$, the streamwise random sweeping velocity fluctuations $\langle v_1^2 \rangle$, the corresponding integral time scale $T_1$, as well as values for the spanwise random sweeping fluctuations $\langle v_2^2 \rangle$ and $T_2$, and the transverse integral length scale $L_2$.
In the following, we discuss how to obtain these parameters from the LES data.

At this stage it is worth recalling that the derivation of the advection model contains a number of simplifying assumptions.
To derive the analytical model, scale separation between the spatially constant random sweeping advection velocity and the advected velocity fluctuations has been assumed. It has been discussed by Kraichnan \cite{Kraichnan}, that even with spatially slowly varying large-scale random sweeping fluctuations, the random sweeping hypothesis as discussed here remains valid. To make it more realistic, we have additionally included a large-scale flow with a finite correlation time in our model. In the application of the model to LES data, however, we do not distinguish anymore between small- and large-scale fluctuations, but rather apply the model to the overall velocity fluctuations.
For the model, we furthermore restrict ourselves to advection in the plane at hub height; any vertical transport of streamwise velocity fluctuations is therefore neglected. Also, effects like large-scale shear which have been discussed, e.g.~in the context of the elliptic \cite{HeZhang2006} and the Mann \cite{Mann1998} models, are currently not taken into account. Additionally, we have assumed a statistically homogeneous flow throughout the derivation in section \ref{sec:Theory}, which is not met in the wind farm application: due to the equidistantly spaced wind turbines, the continuous shift invariance of statistical quantities implied by homogeneity reduces to a discrete one. For averages in spanwise direction, we take these spatial inhomogeneities into account and distinguish two cases in the following: Firstly, we investigate the space-time correlations along the wind turbine columns. The description of velocity correlations along these lines is the primary purpose of the application of the model. In a second case, the lines between the wind turbine columns are investigated for a comparison. 
As a consequence, averages to compute the reference space-time correlations from the LES and the space correlation to feed the model space-time correlation are taken over all streamwise positions and times, whereas the spanwise average is restricted to the corresponding subset of lines. Also the correlation times $T_1$ and $T_2$ are determined for the two cases separately. The distinction in two cases is not done for $U$, $\langle v_1^2 \rangle$, $\langle v_2^2 \rangle$, and $L_2$ as discussed below. In the following, averaging brackets such as $\langle ... \rangle_{\bs x t}$ denote an average over all times and the whole $x_1, x_2$-plane. A subscript $\tilde{\bs x}$ at the $\langle ... \rangle_{ \tilde{\bs x} t}$-brackets denotes an average over all $x_1$ in streamwise direction and a spanwise average over $\tilde{x}_2$ referring to the reduced subset of lines.

\begin{figure}
 \centering
    \includegraphics[trim =  0.5cm 0.5cm  2.5cm 1.8cm, clip, width=0.49\textwidth]{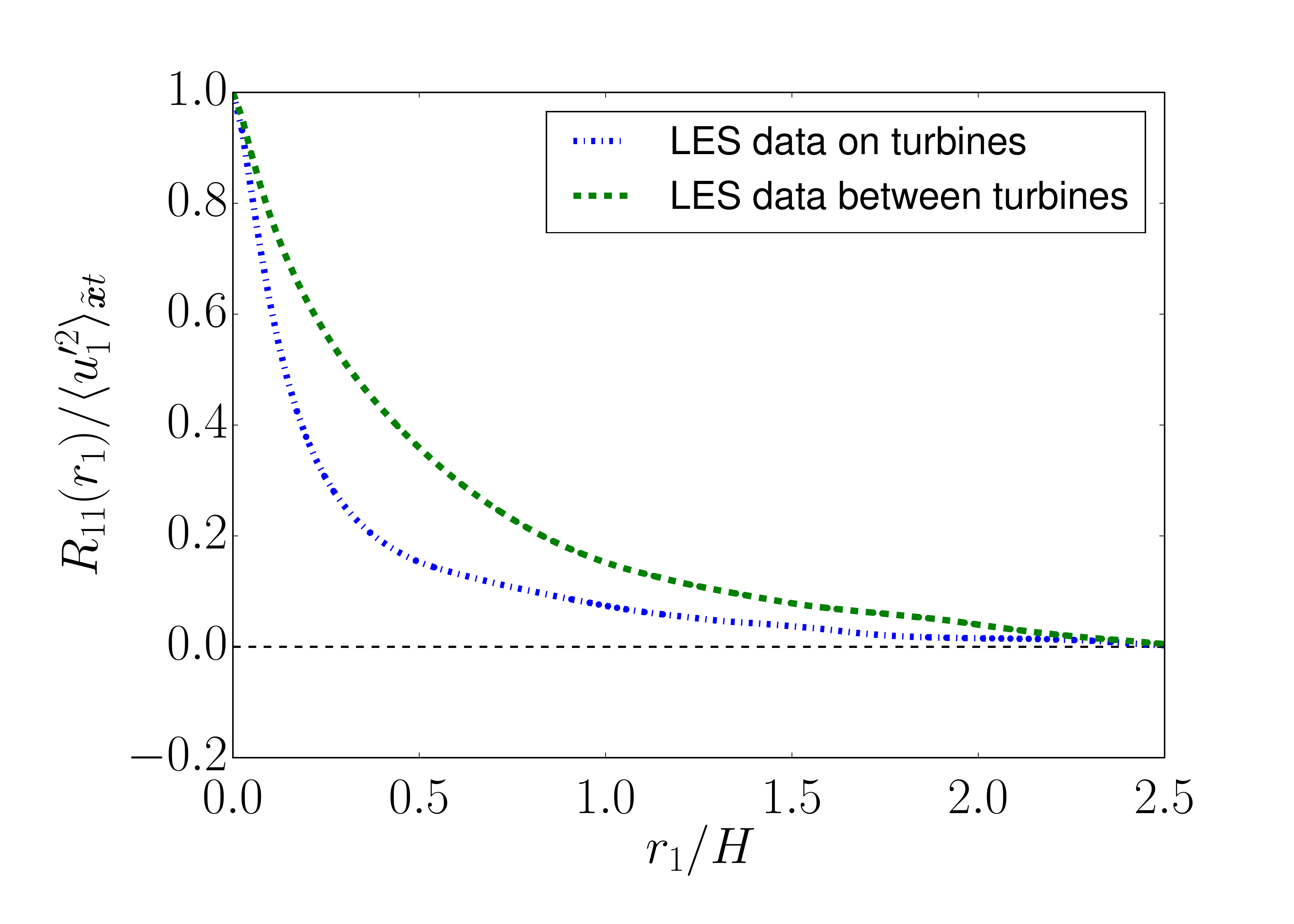}
    \includegraphics[trim =  0.5cm 0.5cm  2.5cm 1.8cm, clip, width=0.49\textwidth] {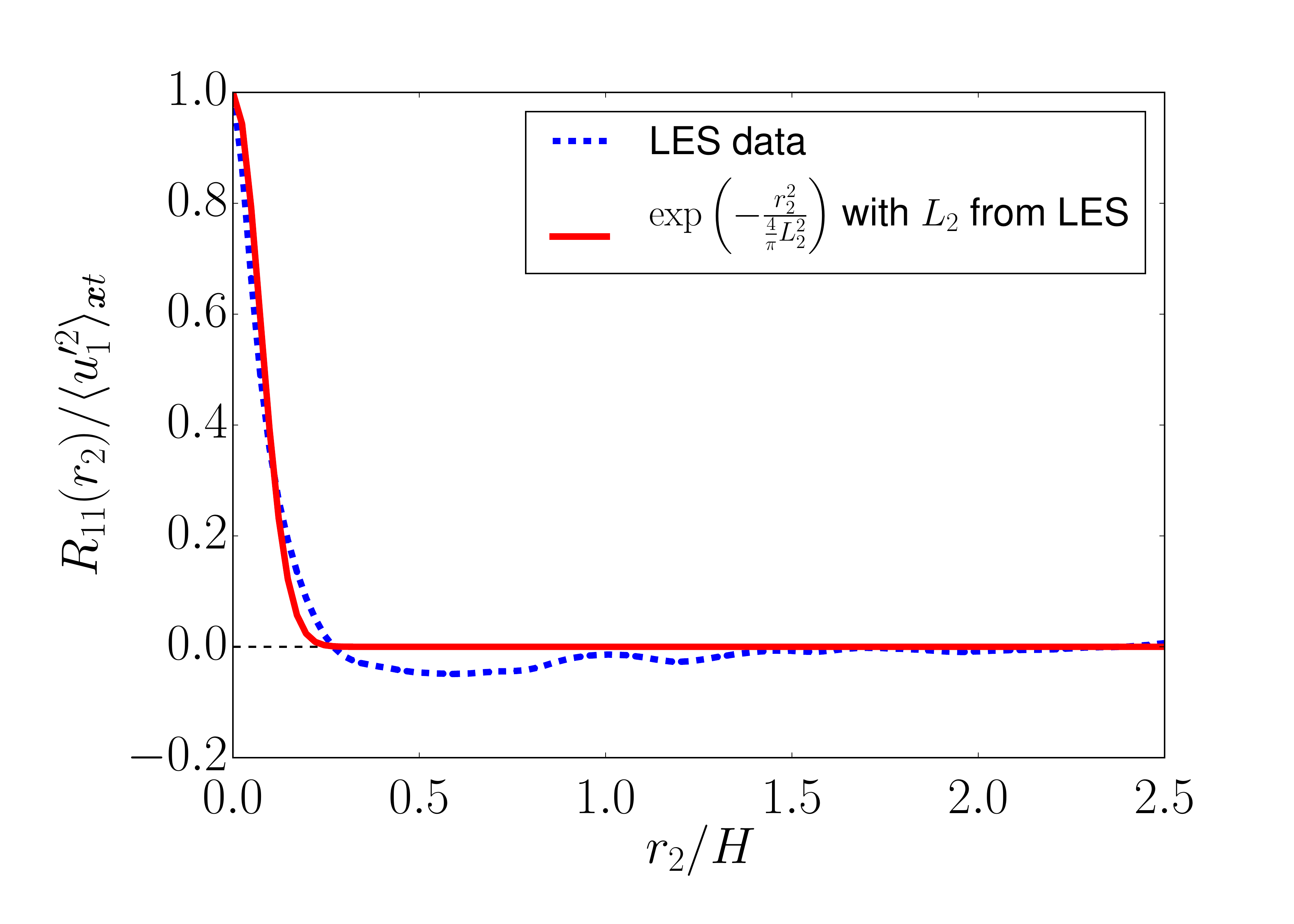}
    \caption{
    Left: Comparison of the longitudinal normalized streamwise velocity space correlations obtained from LES data for the two cases, $R_{11}(r_1 ) = \langle u_1'(x_1 + r_1,x_2 ,t) u_1'(x_1,x_2,t) \rangle_{\tilde{\bs x} t}$. The decay of the correlation is higher on the lines of the turbines. 
    Right: transverse normalized spatial correlation of the streamwise velocity fluctuation,
$R_{11}(r_2 ) = \langle u_1'(x_1,x_2+r_2,t) u_1'(x_1,x_2,t) \rangle_{\bs x t}$. The average here is taken over the whole $x_2$-range. The integral over the positively correlated range of the blue dashed line (LES) is used to determine $L_2$, cf.~equation \ref{eq:L2_integral}. The transverse velocity correlation is modeled as a Gaussian (red solid line) in the modeling process in section \ref{subsec:Reduction}.}
    \label{fig:figure5_6}
  \end{figure}

Figure \ref{fig:figure5_6} shows the longitudinal and transverse spatial velocity correlations. The longitudinal correlations for the two cases are shown in the left plot. For a comparison, the right panel of figure \ref{fig:figure5_6} shows the transverse space correlation which is used for the computation of the integral length scale $L_2$ below.

As for the mean velocity $U$, we use the overall mean velocity, i.e.~the temporally and spatially averaged streamwise velocity $\langle u_1 \rangle_{\bs x t} = 7.88\, u^*$ for both cases. However, as a comparison, the mean velocity along the lines of the turbines is $\langle u_1 \rangle_{\tilde{\bs x} t} = 6.51\, u^*$, while it is $\langle u_1 \rangle_{\tilde{\bs x} t} = 8.59\, u^*$ over the subset of lines between the turbines.

The second moment of the streamwise and spanwise random sweeping fluctuations are obtained from the LES straightforwardly by computing the variance of the respective velocity components, i.e.~$ \langle v_1^2 \rangle = \langle u_1^2 \rangle_{\bs x t}-\langle u_1 \rangle_{\bs x t}^2 =  3.60\, u^{*2}$, and $ \langle v_2^2 \rangle = \langle u_2^2 \rangle_{\bs x t}=  1.48\, u^{*2}$ (note that there is no imposed mean flow in the spanwise direction). The way $\langle v_1^2 \rangle$ is computed does not yield the same result as a spatial average of equation \eqref{eq:time_average}, i.e.~$\langle u_1'^2 \rangle_{\bs x t}= 2.68\, u^{*2}$.
For $\langle v_1 \rangle$ we do not distinguish the two cases because the highest turbulence intensity is at the edges of the wakes behind the turbine, cf.~figure \ref{fig:figure3_4} which is neither exactly on the lines of the turbines nor exactly between the turbines. For the spanwise random sweeping fluctuations we also use the overall mean square spanwise fluctuations from LES.

\begin{figure}
 \centering
    \includegraphics[trim =  0.5cm 0.5cm  2.5cm 1.8cm, clip, width=0.49\textwidth]{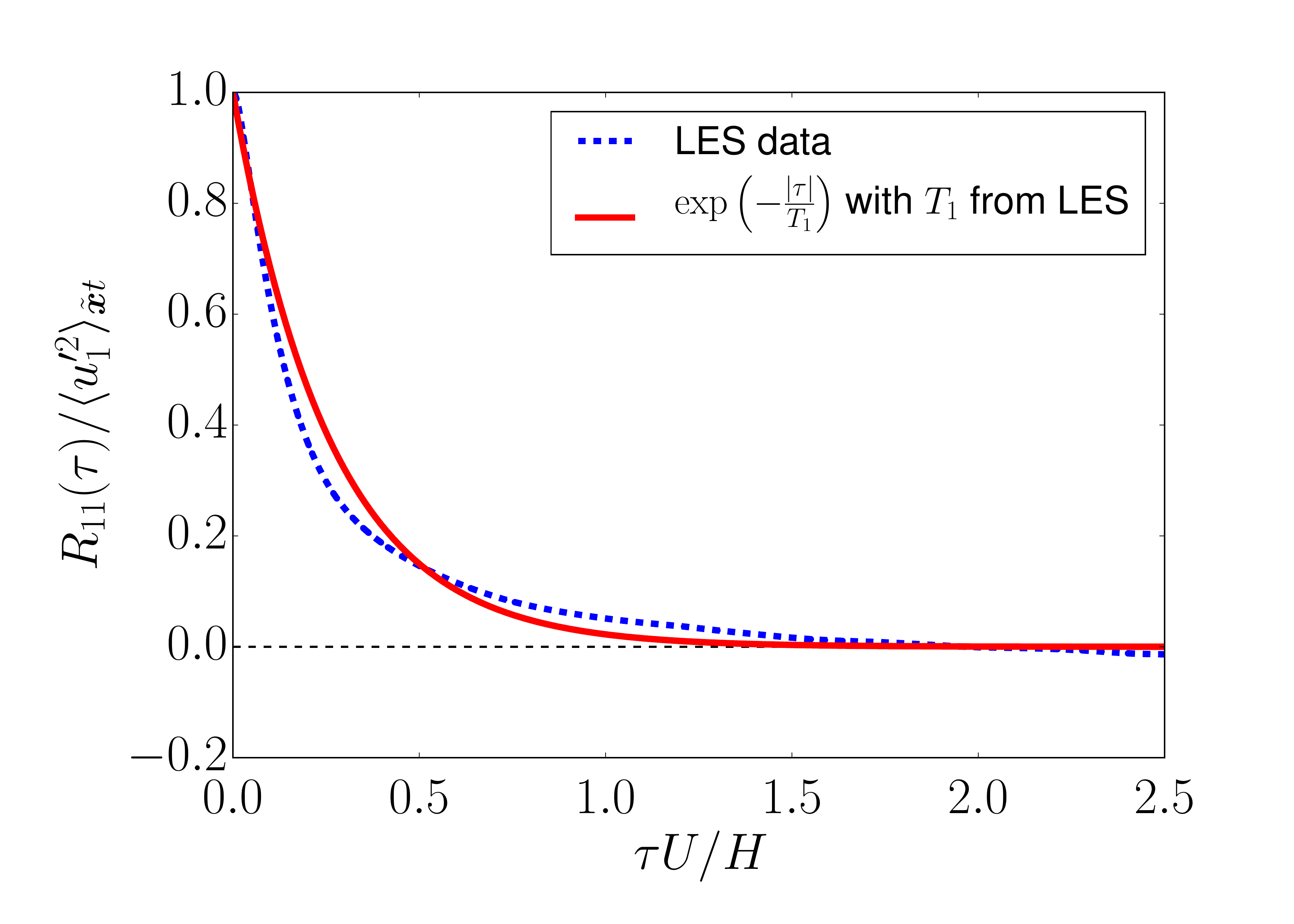}
    \includegraphics[trim = 0.5cm 0.5cm  2.5cm 1.8cm, clip, width=0.49\textwidth]{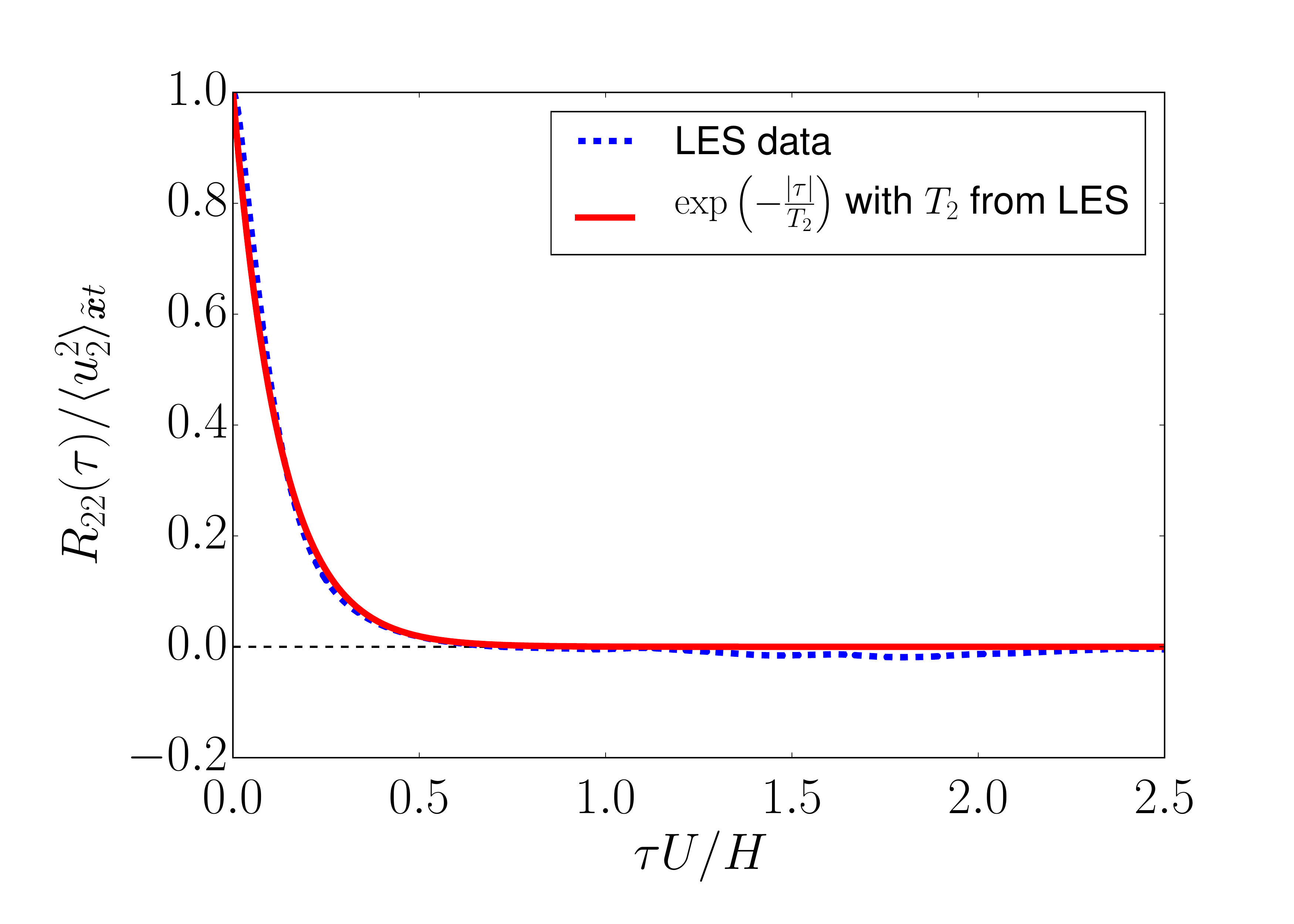}
    \caption{
    Comparison of the normalized temporal velocity correlations on the turbine columns from LES data (blue dashed line) to the exponential decorrelation in equation \eqref{eq:exp_decay_rsvel}, (red solid line). Left: normalized streamwise temporal velocity correlation $R_{11}( \tau)= \langle u_1'(\bs x ,t + \tau) u_1'(\bs x,t) \rangle_{\tilde{\bs x} t}$ from LES data and exponential decorrelation with the same streamwise integral time scale as given through the LES. 
    Right: normalized spanwise temporal velocity correlation $R_{22}(\tau)= \langle u_2(\bs x ,t + \tau) u_2(\bs x,t) \rangle_{\tilde{\bs x} t}$ from LES data and exponential decorrelation with the same spanwise integral time scale as given through the LES.}
    \label{fig:figure7_8}
  \end{figure}

\begin{figure}
    \includegraphics[trim =  0.5cm 0.8cm 3.5cm 1.8cm, clip, width=0.49\textwidth]{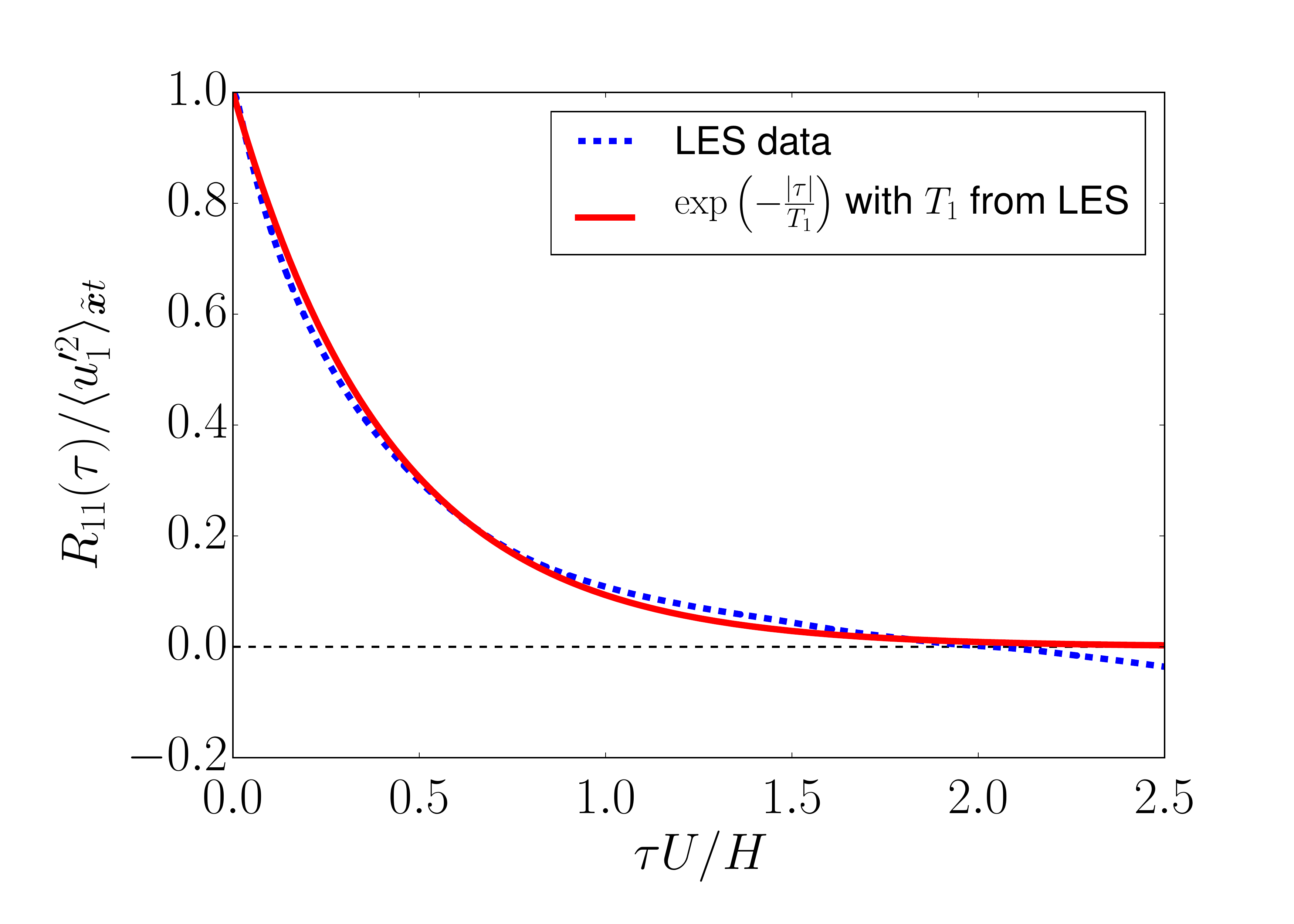}
    \includegraphics[trim =  0.5cm 0.8cm 3.5cm 1.8cm, clip, width=0.49\textwidth]{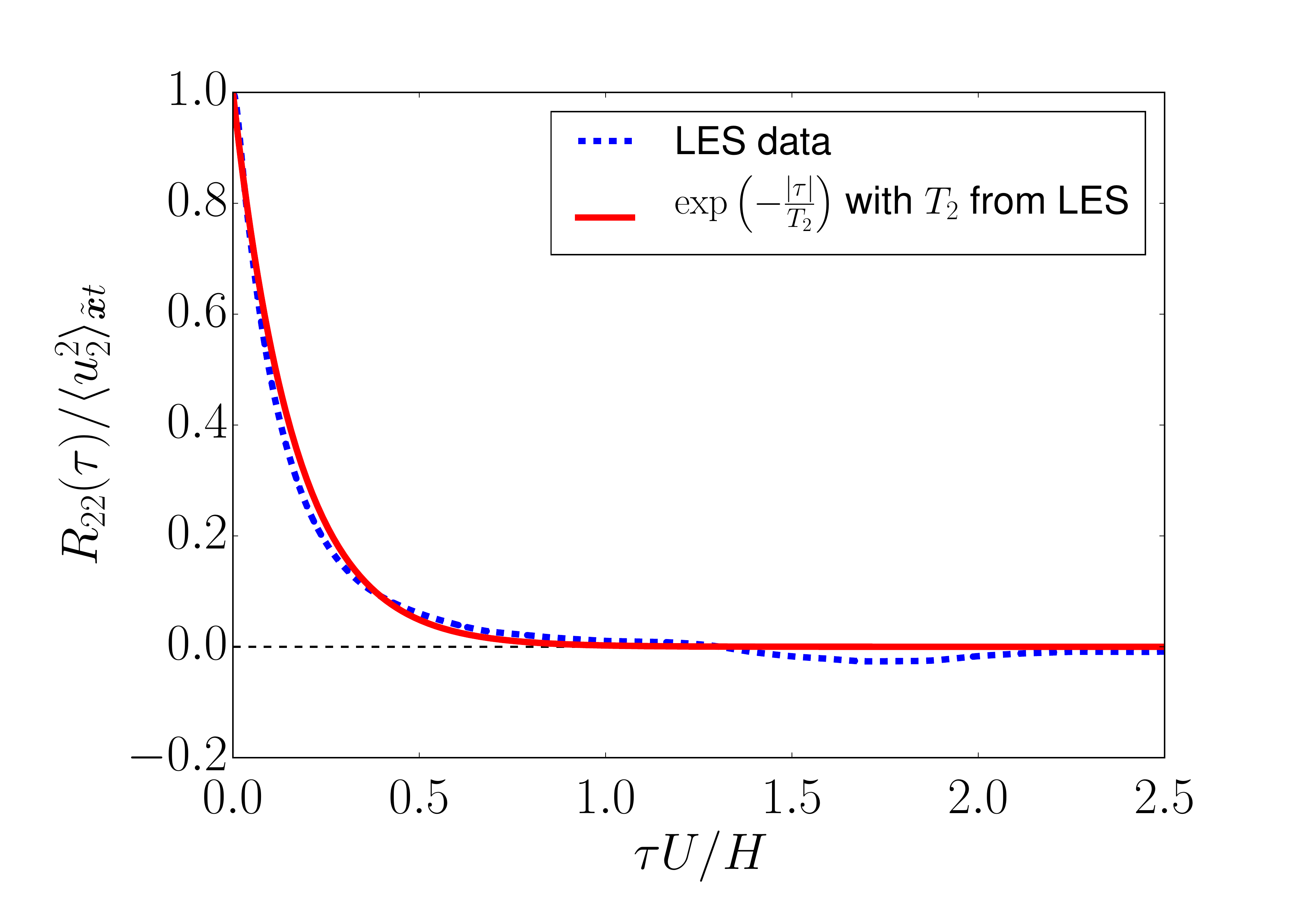}
    \caption{Comparison of the normalized temporal velocity correlations between the turbine columns from LES data (blue dashed line) to the exponential decorrelation in equation \eqref{eq:exp_decay_rsvel}, (red solid line). Left: normalized streamwise temporal velocity correlation $R_{11}(\tau)$ from LES data and exponential decorrelation with the same streamwise integral time scale as given through the LES. 
    Right: normalized spanwise temporal velocity correlation $R_{22}(\tau)$ from LES data and exponential decorrelation with the same spanwise integral time scale as given through the LES.}
    \label{fig:figure9_10}
  \end{figure}

We distinguish the two cases for the determination of the correlation time scales of the random sweeping velocities. The model time scales $T_1$ and $T_2$ are determined from the temporal correlation of the LES streamwise and spanwise velocity fluctuations with respect to the subset of lines in the respective cases. The integration stops at the first zero crossing:
\begin{equation}
\label{eq:T1_T2}
 T_1  = \frac{1}{\langle u_1'^2\rangle_{\tilde{\bs x} t}} \int\limits_{0}^{\tilde{t}} \mathrm{d}\tau \, \langle u_1'(\bs x , t+\tau) u_1'(\bs x,t)\rangle_{\tilde{\bs x} t}  \, , \ \ 
 T_2   = \frac{1}{\langle u_2^2\rangle_{\tilde{\bs x} t}} \int\limits_{0}^{\tilde{t}} \mathrm{d}\tau \, \langle u_2(\bs x , t+\tau) u_2(\bs x,t)\rangle_{\tilde{\bs x} t}  \, .
\end{equation}
The $\tilde{t}$ indicates the first zero-crossing of the correlation so that for the integral time scales only the positively correlated range is counted. 
As a result, the model integral time scales $T_1$ and $T_2$ on the lines of turbines (case 1) are $T_1 = 0.033\, H/u^*$ and $T_2 = 0.016\, H/u^*$. The time scales between the lines of the turbines (case 2) result in $T_1 = 0.054\, H/u^*$ and $T_{2} = 0.021  \, H/u^*$.
The dashed blue lines in figures \ref{fig:figure7_8} and \ref{fig:figure9_10} are the normalized streamwise and the spanwise temporal correlation from simulation data. The solid red lines are exponentials with the above given integral time scales which compare favorably to the LES data.
As a reminder, we also use the LES data for obtaining the streamwise space correlation $R_{11}(r_1)$ for the model. The transverse correlation $R_{11}(r_2)$ is modeled as a Gaussian in section \ref{subsec:Reduction} which requires the determination of the transverse integral length scale $L_2$.
The transverse integral length scale $L_2$ of the streamwise velocity is taken over the whole spanwise direction, without a distinction of the two cases, by
\begin{equation}
  L_2 =  \frac{1}{\langle u_1'^2\rangle_{\bs x t}}\int\limits_{0}^{\tilde{l}} \mathrm{d}r_2 \, \langle u_1'(x_1, x_2 +r_2, t) u_1'(x_1,x_2,t)\rangle_{\bs x t}\, .
  \label{eq:L2_integral}
\end{equation}
As above, the $\tilde{l}$ indicates the first zero-crossing of the correlation so that also for the integral length scale only the positively correlated range is taken into account. The right-hand plot of figure \ref{fig:figure5_6} shows the transverse spatial correlation of the streamwise velocity component from the LES in comparison to the Gaussian model. The choice of the Gaussian with the integral length scale $L_2$ as computed by \eqref{eq:L2_integral} shows good agreement with the LES data.

A summary of the free parameters is shown in table \ref{tab:parameters}. In fact, the free parameters from the case along the lines of turbines and between differ essentially only in the computation of the space correlation data along the respective subset of lines and the integral time scales $T_1$ and $T_2$. With the possibility to set the free parameters, the characteristics of the LES space-time correlation can be captured, e.g.~the choice of the spanwise random sweeping variance and the integral length scale influences the amplitude of the time kernel, whereby the streamwise random sweeping influences the amplitude and the width.

\begin{table}
\begin{center}
\begin{tabular}{l| p{3cm} |p{3cm}  }
   Parameter evaluated from LES & {Case 1: along\newline the lines of turbines} & Case 2: between\newline the lines of the turbines \\
  \hline
 $\phantom{\langle} U = \langle u_1 \rangle_{\bs x t}$  & \multicolumn{2}{c}{ $ 7.88\, u^{*\phantom{2}}$ } \\
$ \langle v_1^2 \rangle =   \langle u_1^2 \rangle_{\bs x t} -\langle u_1 \rangle^2_{\bs x t} $  & \multicolumn{2}{c}{$3.60 \, u^{*2}$ } \\
$ \langle v_2^2 \rangle =   \langle u_2^2 \rangle_{\bs x t}  $  & \multicolumn{2}{c}{$1.48 \, u^{*2}$ } \\
$ \phantom{\langle} L_2  $  &  \multicolumn{2}{c}{$0.09 \, H\phantom{^{*2}}$}  	 \\
 \hline
$\phantom{\langle} T_1  $     & $0.033 \, H/u^*$ &  $0.054 \, H/u^*$ \\
$\phantom{\langle} T_2 $   & $0.016 \, H/u^*$ &  $0.021  \, H/u^*$\\
 \end{tabular}\\
  \end{center}
    \caption{Free parameters for the model for the two cases; $U$, $\langle v_1^2\rangle$,  $\langle v_2^2\rangle$, and $L_2$ are the same for both cases whereas the integral time scales change. Furthermore, the space correlation $R_{11}(r_1)$ which is used as input for the model differs for the two cases. 
    }
  \label{tab:parameters}
  \end{table}

\section{Comparison of the space-time correlation model with LES data}
\label{sec:Comparison}

For the comparison of our model space-time correlations to LES, we consider velocity correlations both on wind turbines and between turbine columns in the following sections \ref{subsec:on turbines} and \ref{subsec:between turbines} separately. In section \ref{subsec:discussion}, the results are discussed and the two cases are compared.

\subsection{Space-time correlations on wind turbine columns}
\label{subsec:on turbines}
In the first case, all spanwise averages for the LES space-time correlation $R_{11}(r_1, \tau)$ and for the space correlation $R_{11}(r_1)$ are taken with respect to the reduced subset of $\tilde{x}_2$-coordinates along the lines of turbines, i.e.~the spanwise average over 16 lines separated by the spanwise turbine spacing.
The parameters $U$, $\langle v_1^2 \rangle$, $\langle v_2^2\rangle$, $T_1$, $T_2$, and $L_2$ are chosen according to case 1 in table \ref{tab:parameters}.
\begin{figure}
  \centering
    \includegraphics[trim = 0cm 0cm 0cm 0cm, clip,  width=0.49\textwidth]{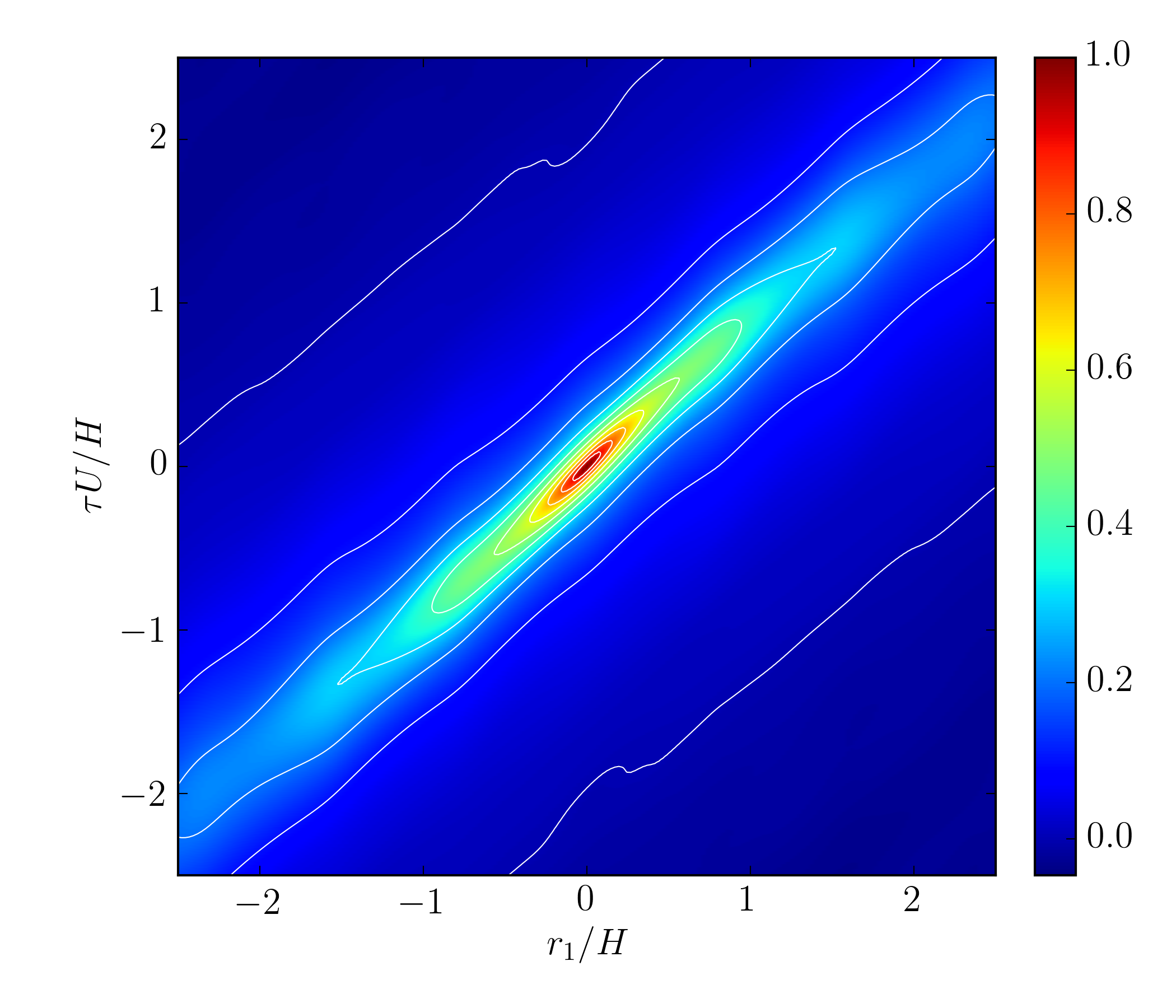}
    \includegraphics[trim = 0.0cm 0cm 0cm 0cm, clip,  width=0.49\textwidth]{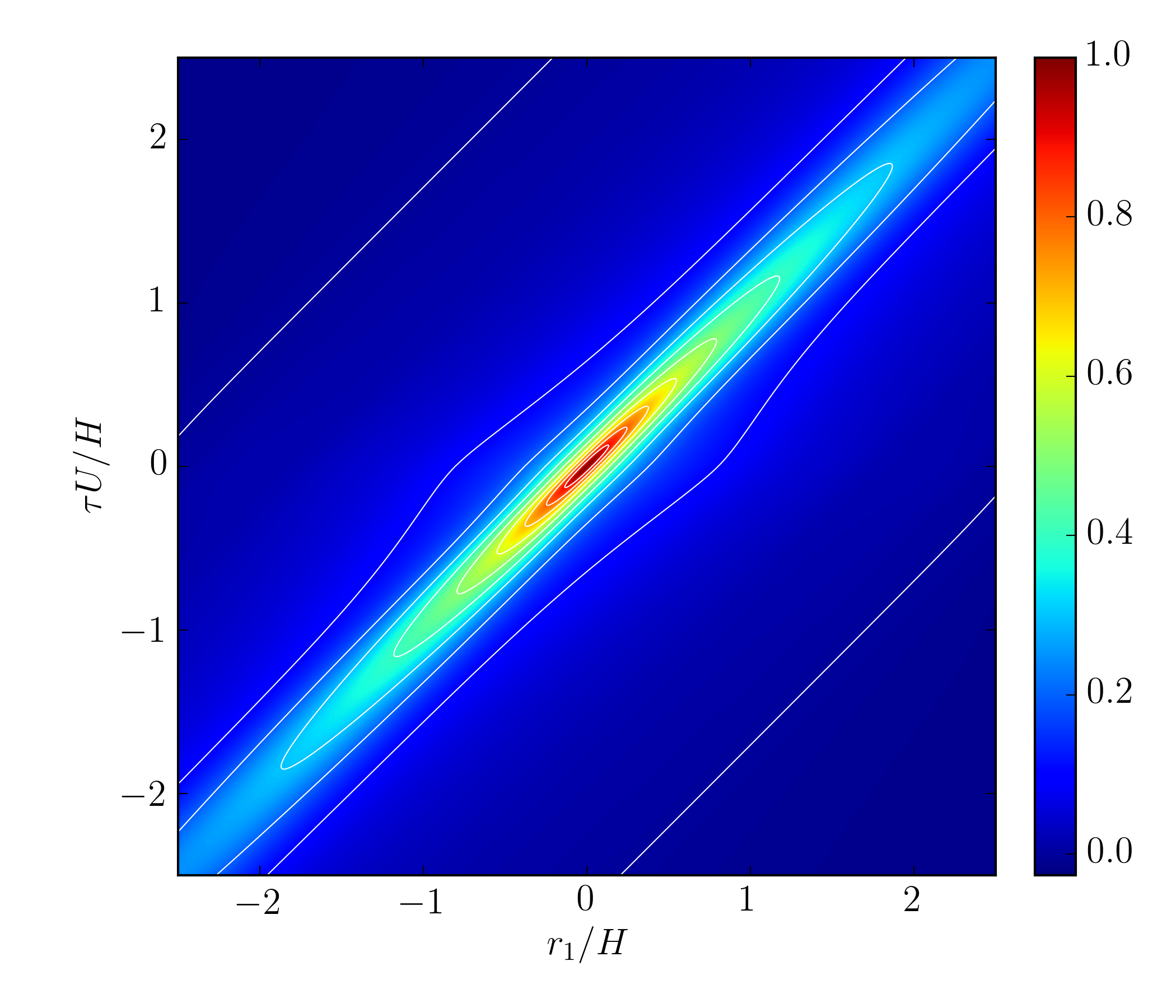}
    \caption{Comparison of space-time correlations $R_{11}(r_1, \tau)/\langle u_1'^2 \rangle_{\tilde{\bs x} t}$ on the turbine columns from LES data to the model space-time correlations. 
    Left: normalized space-time correlation of the streamwise velocity component from LES data on lines of wind turbines. White isocontours are shown for the interval $[0.0, 1.0]$ in $0.1$-steps. The slope of the LES space-time correlation originates from the choice of $U$ which is used to non-dimensionalize $\tau$, a slightly larger $U$ would increase the slope towards 45 degrees. Right: normalized model space-time correlation with $R_{11}(r_1)$ from LES with parameters $U$, $\langle v_1^2 \rangle$, $\langle v_2^2 \rangle$, $T_1$, $T_2$, $L_2$ as given in table \ref{tab:parameters} (case 1). 
    The kink in the isocontour lines in the right-hand side plot results from the analytical form of the decorrelation kernel.}
    \label{fig:figure11_12}
  \end{figure}

The contour plots of the LES space-time correlation and the model space-time correlation are shown in figure \ref{fig:figure11_12} which allow for a qualitative comparison.
The left panel of figure \ref{fig:figure11_12} shows the space-time correlation evaluated from the LES data. The streamwise velocity fluctuations decorrelate both in space as well as in time. The characteristic tilt of the space-time correlation is clearly given by the mean flow advection. The highly correlated region (red color) is small, the main decorrelation takes place within one turbine spacing, $|r_1|/H \leq 0.785$ and one inter-turbine travel time (i.e.~the time for traveling from one turbine to the next, here with the mean velocity $U = 7.88 \, u^*$), $|\tau| U/H \leq 0.785$.
In this iso-correlation contour plot, we find wiggles caused by the periodic modulation of the flow due to the presence of wind turbines. The right panel of figure \ref{fig:figure11_12} shows the model space-time correlation where the free parameters have been obtained from the LES which clearly shows the same characteristics as the space-time correlations from LES in the left panel. The wiggles are not captured by the model. 

\begin{figure}
  \centering
    \includegraphics[trim = 0.5cm 0.5cm 2.5cm 1.8cm, clip,  width=0.49\textwidth]
    {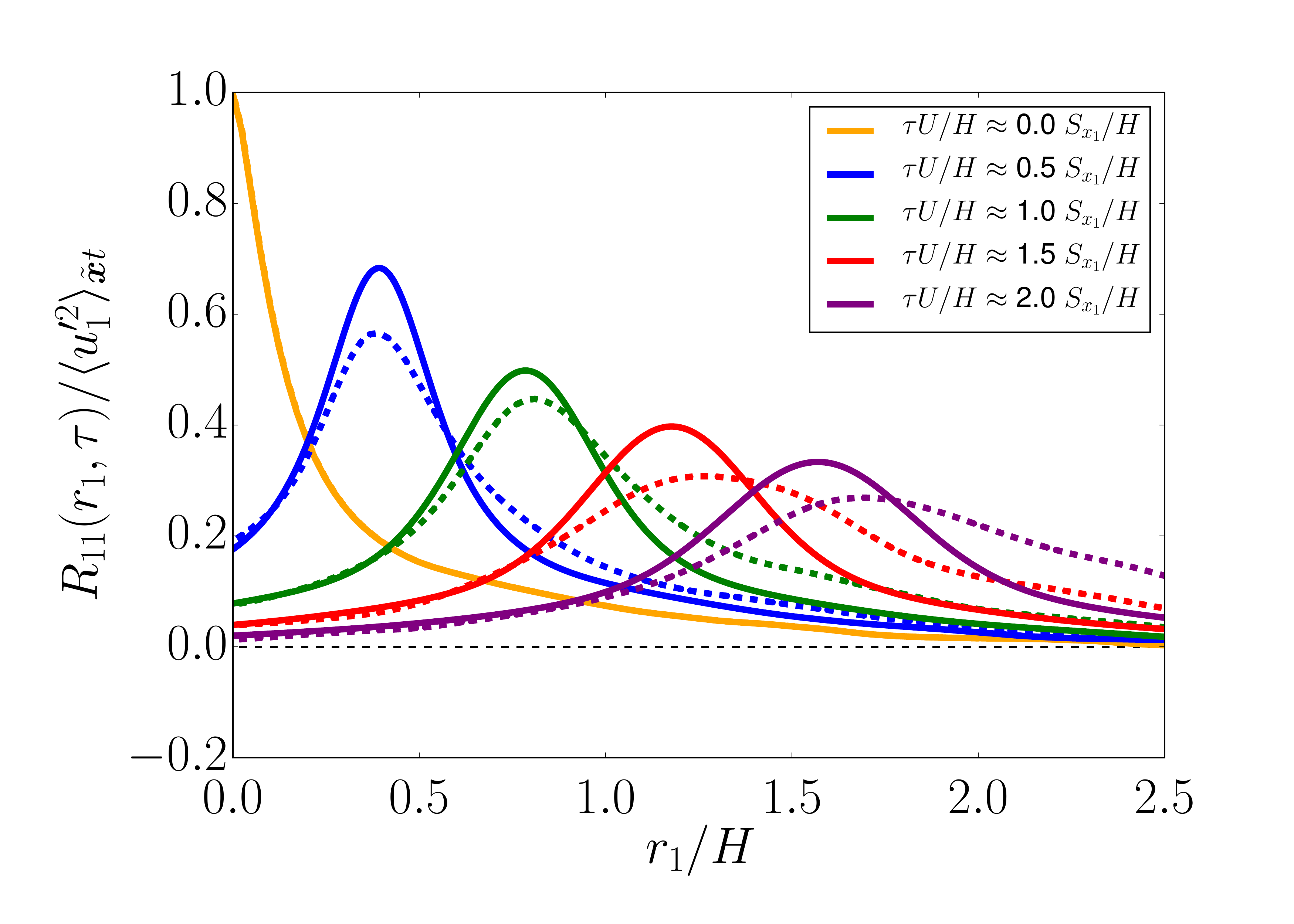}
  \centering
    \includegraphics[trim = 0.5cm 0.5cm 2.5cm 1.8cm, clip,  width=0.49\textwidth]{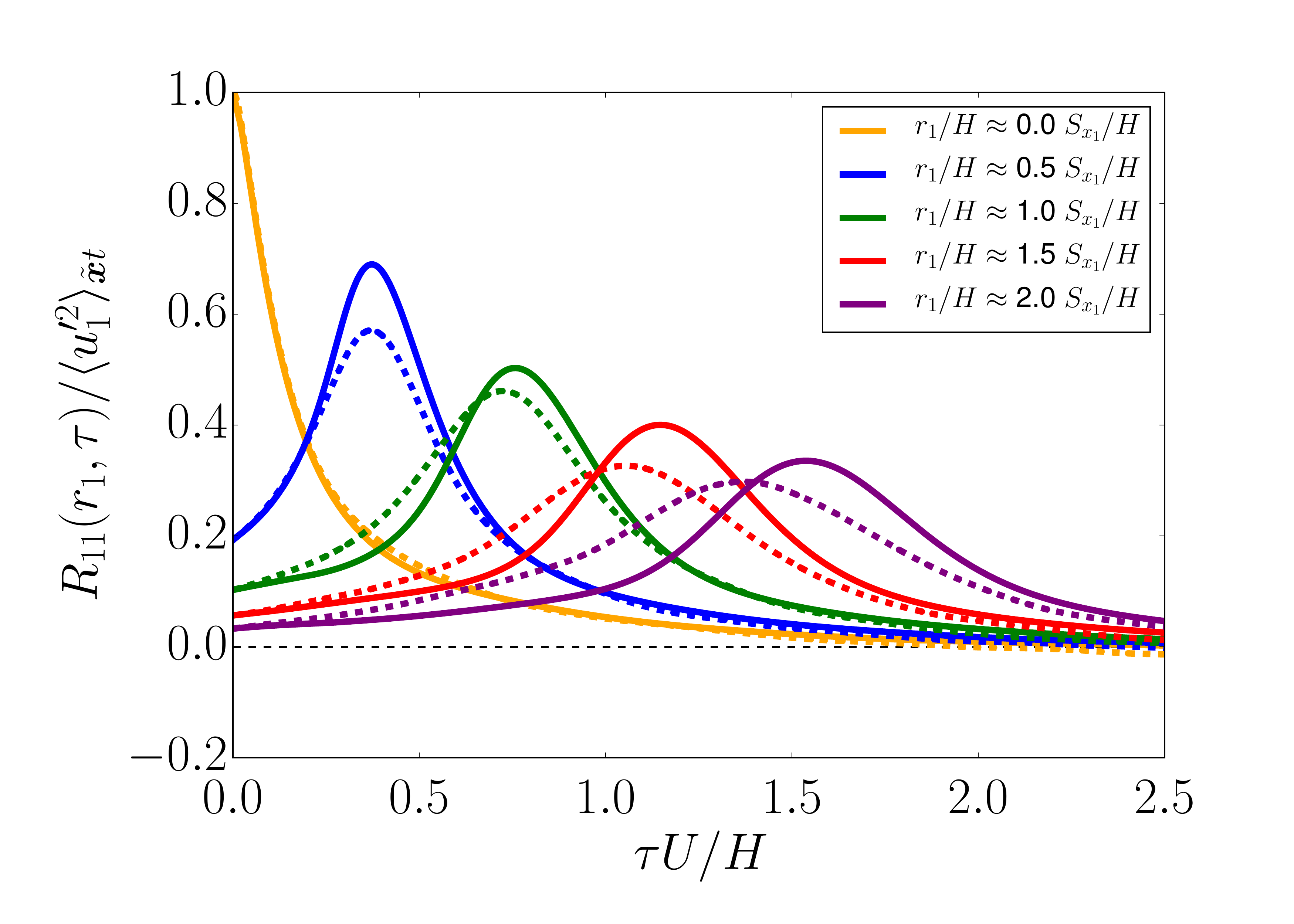}
\caption{Horizontal and vertical cuts through the contour plots \ref{fig:figure11_12} for the comparison of space-time correlations on the turbine columns from LES to model space-time correlations. Dashed lines correspond to the LES data, solid lines to the model space-time correlation with $R_{11}(r_1)/\langle u_1'^2 \rangle_{\tilde{\bs x} t}$ from LES, and $ U$, $\langle v_1^2 \rangle$, $\langle v_2^2 \rangle$, $T_1$, $T_2 $, $L_2$ as given in case 1 in table \ref{tab:parameters}. Left: normalized $R_{11}(r_1, \tau)$ at fixed instants in time as a function of $r_1$. The solid orange curve in the left panel is per definition the normalized space correlation $R_{11}(r_1)/\langle u_1'^2 \rangle_{\tilde{\bs x} t}$ from LES data. Right: normalized $R_{11}(r_1, \tau)$ at fixed spatial increments. $S_x=0.785\, H$ is the turbine spacing. 
The dashed orange curve from the LES in the right panel for $r_1/H = 0$ is the same as the dashed blue line in the left panel in figure \ref{fig:figure7_8}. Although not exactly the same, the orange solid model curve for $r_1/H = 0$ is in good approximation the corresponding exponential in the left panel of figure \ref{fig:figure7_8}.}
\label{fig:figure13_14}
\end{figure}

For a more quantitative comparison, figure \ref{fig:figure13_14} shows cuts through the contour plots in figure \ref{fig:figure11_12}. The cuts through the contour plots confirm a very good agreement with the LES data, demonstrating that our model for the space-time correlations captures the temporal decorrelation trends well. The temporal increments in the left panel of figure \ref{fig:figure13_14} are given in terms of the inter-turbine travel times. The spatial increments in the right panel are related to the turbine distance. The results are comparable also for other increment sizes than the ones chosen in the right panel of figure \ref{fig:figure13_14}, as can be confirmed through figure \ref{fig:figure11_12}.

The  value of the overall mean velocity $U = 7.88 \, u^*$ (cf.~table \ref{tab:parameters}) matches well with the moving peaks. 
The amplitude and width of the curves corresponding to full multiple turbine distances or full multiple inter-turbine travel times show a visibly better agreement between the model and the LES. 
Furthermore, the amplitude of the LES correlation does not seem to change considerably when increasing the spatial increment from half increments of turbine distances to full integer multiples of turbine distances. 
This effect causes the wiggles in the LES contour plot \ref{fig:figure11_12} are not directly visible in the cuts in figure \ref{fig:figure13_14}. They can be observed indirectly by the fact that the amplitude of the curves stay constant over several increment sizes.
This may be related to the fact that the main decorrelation takes place within half the turbine distance to the downstream turbine. The turbulence intensity is considerably higher in the first half distance and then decreases immediately in front of the next turbine, cf.~right panel in figure \ref{fig:figure3_4}.

\subsection{Space-time correlations between wind turbine columns}
\label{subsec:between turbines}
In the second case, $R_{11}(r_1, \tau)$ and $R_{11}(r_1)$ from LES are computed with respect to the 16 lines between the turbines, separated by the spanwise turbine spacing. The values for $U$, $\langle v_1^2 \rangle$, $\langle v_2^2 \rangle$, $T_1$, $T_2$, and $L_2$ are chosen according to case 2 in table \ref{tab:parameters}.

\begin{figure}
  \centering
    \includegraphics[trim = 0cm 0cm 0cm 0cm, clip,  width=0.49\textwidth]{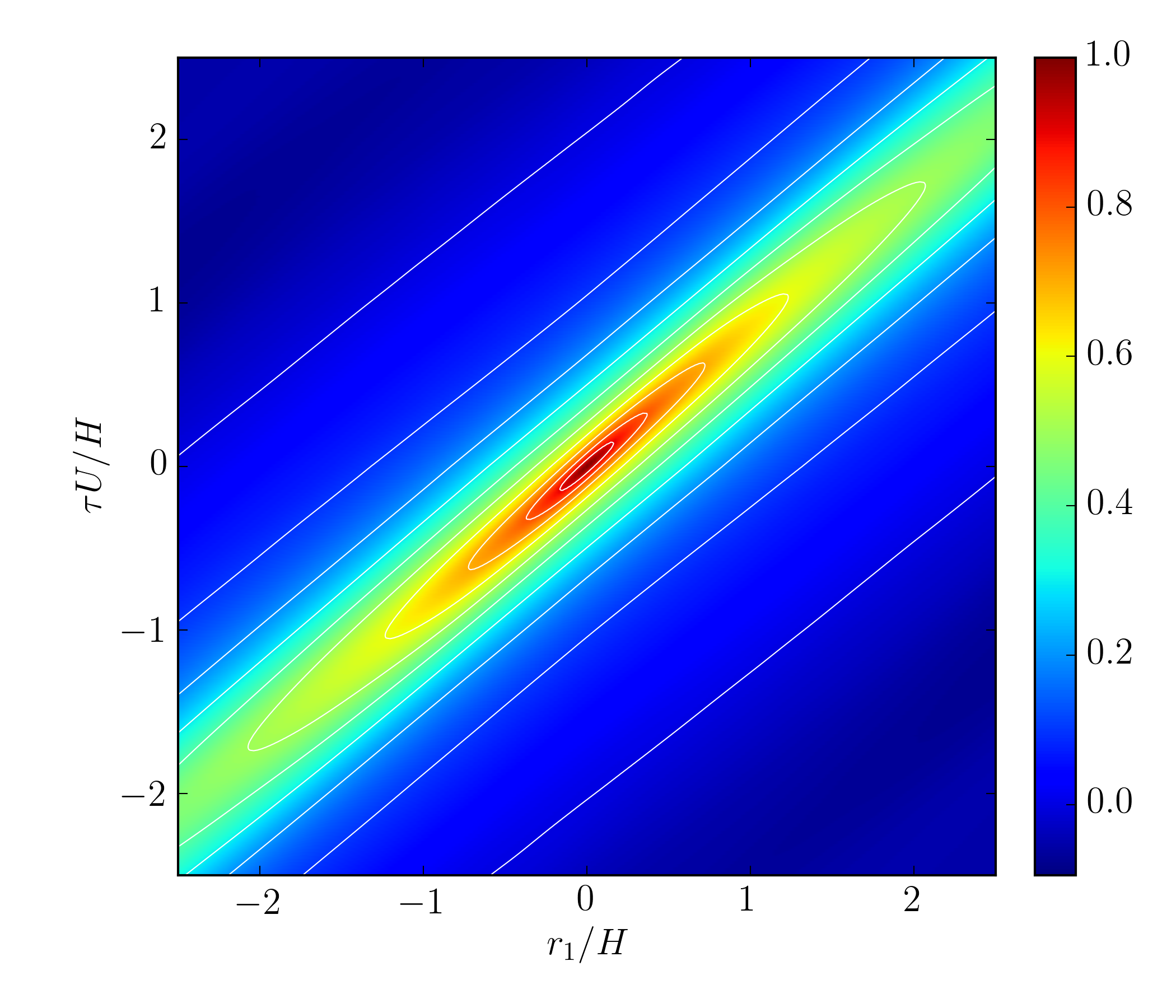}
    \includegraphics[trim = 0.0cm 0cm 0cm 0cm, clip,  width=0.49\textwidth]{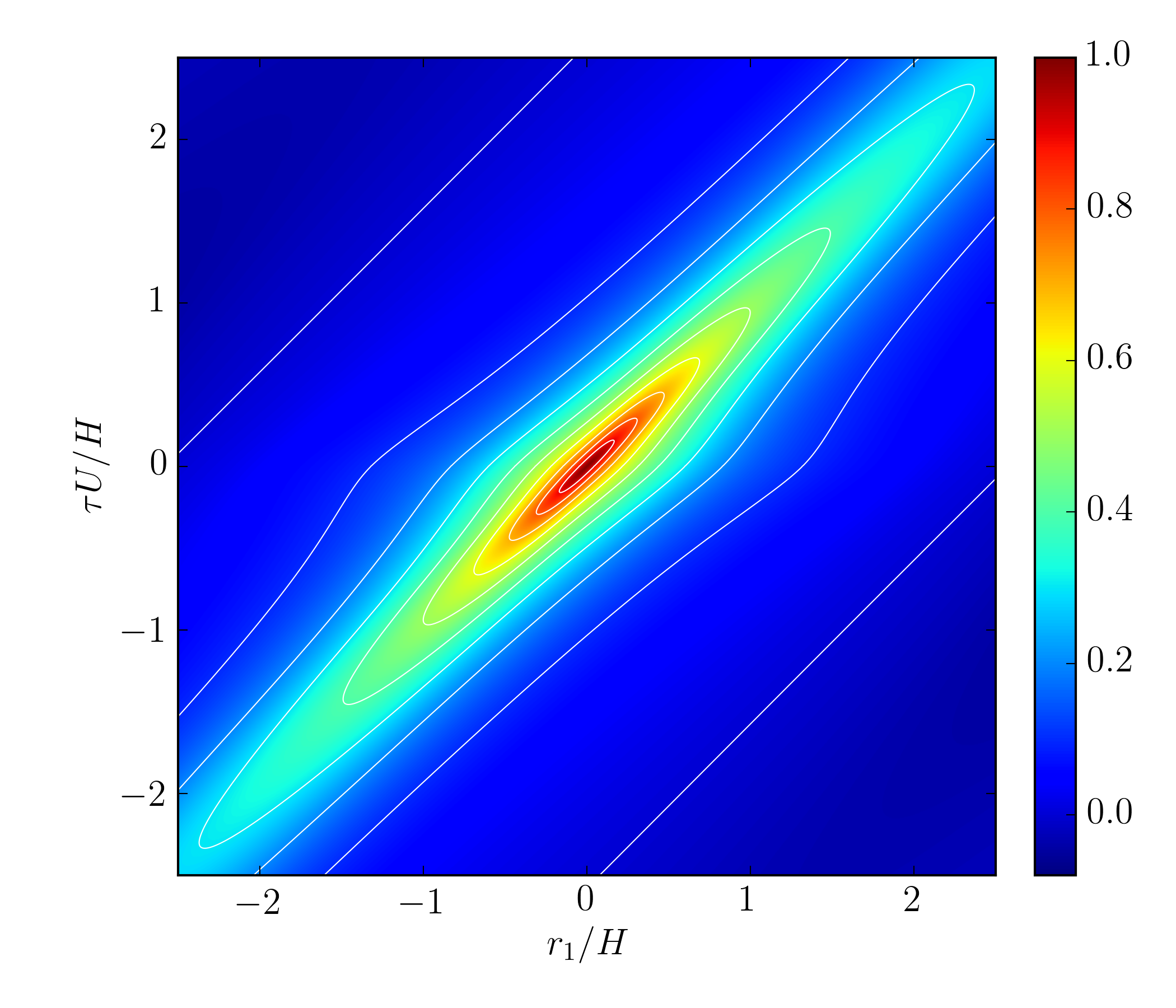}
    \caption{Comparison of space-time correlations $R_{11}(r_1, \tau)/\langle u_1'^2 \rangle_{\tilde{\bs x} t}$ between the turbine columns from LES data to the model space-time correlations. Left: normalized space-time correlation of the streamwise velocity component from LES data. White isocontours are shown for the interval $[0.0, 1.0]$ in $0.1$-steps. The slope of the LES space-time correlation can be increased by increasing $U$. Right: normalized model space-time correlation with $R_{11}(r_1)$ from LES and parameters $U$, $\langle v_1^2 \rangle$, $\langle v_2^2 \rangle $, $T_1$, $T_2$, $L_2$ according to table \ref{tab:parameters}, case 2. As in the previous case along the turbine lines, the kink in the isocontour lines of the model plot results from the decorrelation kernel.
    }
    \label{fig:figure15_16}
  \end{figure}

\begin{figure}
  \centering
    \includegraphics[trim = 0.5cm 0.5cm 2.5cm 1.8cm, clip,  width=0.48\textwidth]{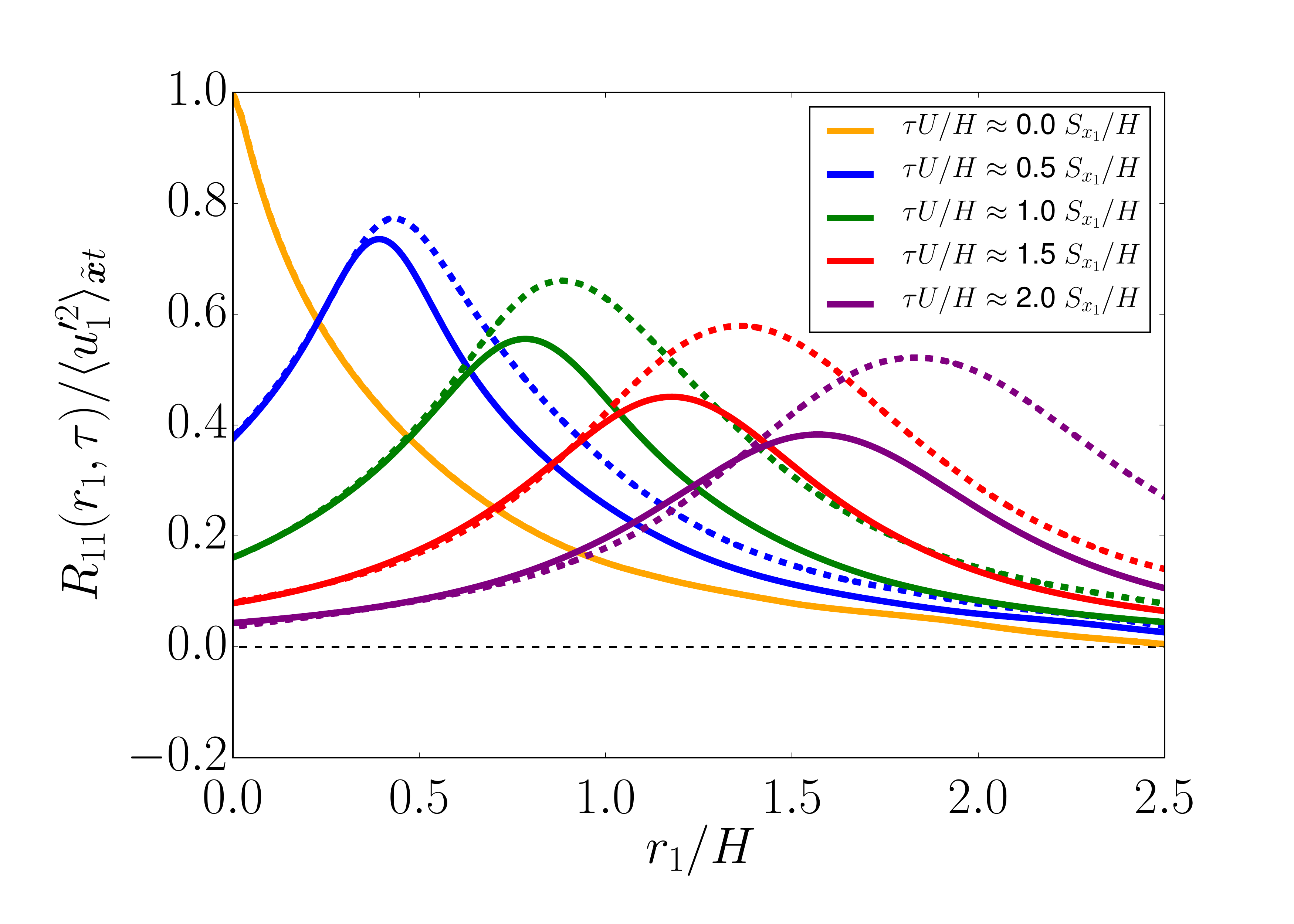}
    \includegraphics[trim = 0.5cm 0.5cm 2.5cm 1.8cm, clip,  width=0.48\textwidth]{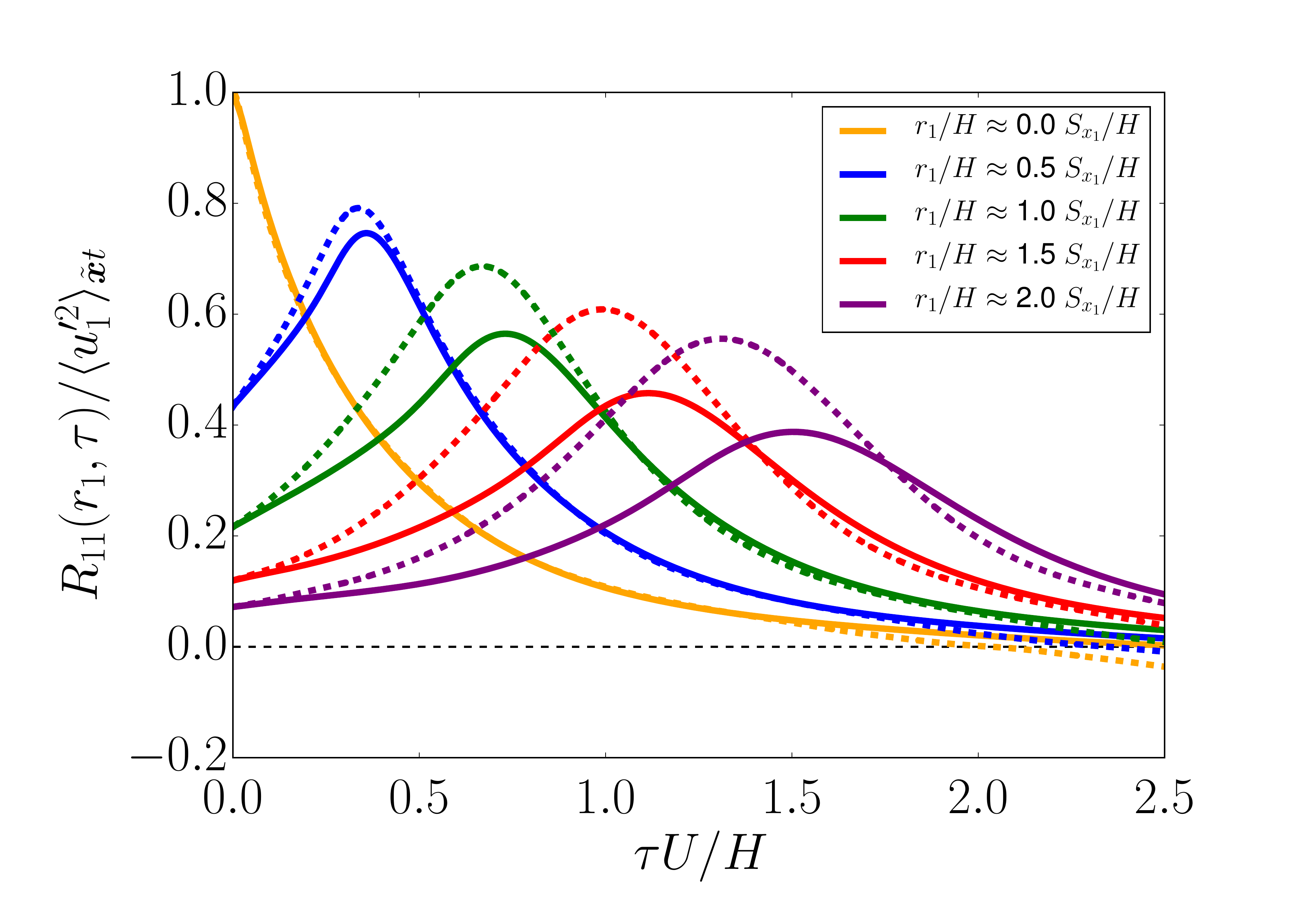}
   \caption{Horizontal and vertical cuts through the contour plots \ref{fig:figure15_16} for the comparison of  normalized space-time correlations between the turbine columns from LES to the model space-time correlations.
   Dashed lines correspond to the LES data, solid lines to the normalized model space-time correlation with $R_{11}(r_1)/\langle u_1'^2 \rangle_{\tilde{\bs x} t}$ from LES, and the parameters $ U$, $\langle v_1^2 \rangle$, $\langle v_2^2 \rangle $, $T_1$, $T_2$, $L_2 $ from table \ref{tab:parameters}, case 2. Left: normalized $R_{11}(r_1, \tau)$ at fixed instants in time as a function of $r_1$. The solid orange curve in the left panel is per definition the normalized space correlation $R_{11}(r_1)/\langle u_1'^2 \rangle_{\tilde{\bs x} t}$ from LES data. Right: normalized $R_{11}(r_1, \tau)$ at fixed spatial increments. $S_x=0.785\, H$ is the turbine spacing. The dashed orange curve from the LES in the right panel for $r_1/H = 0$ is the same as the dashed blue line in the left panel in figure \ref{fig:figure9_10}. As in case 1, the solid orange model model curve approximates the corresponding exponential in figure \ref{fig:figure9_10} well.}
    \label{fig:figure17_18}
  \end{figure}

The LES space-time correlation from between the turbines is shown in the iso-correlation contour plot in the left panel of figure \ref{fig:figure15_16}. The right panel of figure \ref{fig:figure15_16} shows a contour plot of the model space-time correlation \eqref{eq:spacetimecorr}. The model space-time correlation exhibits a less extended high correlated (red) area than the LES space-time correlation. For a more quantitative comparison of the model and the LES space-time correlation, figure \ref{fig:figure17_18} shows cuts through the contour plot \ref{fig:figure15_16} for various choices of $r_1$ and $\tau$. The less extended red area of the model carries over to considerably faster decreasing amplitudes of the model curves in comparison to the LES curves in figure \ref{fig:figure17_18}.
  
Figure \ref{fig:figure17_18} shows that, for small temporal and spatial increments, a value of $U = 7.88\, u^*$ seems to fit reasonably in terms of the moving peak of the model space-time correlation. For larger increments, however, a higher value for $U$ approaching the temporal and spatial mean velocity over the lines between the turbines seems more appropriate. The left panel plot of figure \ref{fig:figure17_18} shows very good agreement for small temporal increments (blue curve). Comparably good results are found for the spatial increments up to one turbine distance, $0.785\, H$, in the right panel of figure \ref{fig:figure17_18}. 
For larger temporal and spatial increments, the model still captures the characteristics of the space-time correlations qualitatively, i.e.~in terms of amplitude, the broadening and the peak shift, however, quantitatively there are discrepancies observable.

\begin{figure}
  \centering
    \includegraphics[trim = 0.5cm 0.5cm 2.5cm 1.8cm, clip,  width=0.48\textwidth]{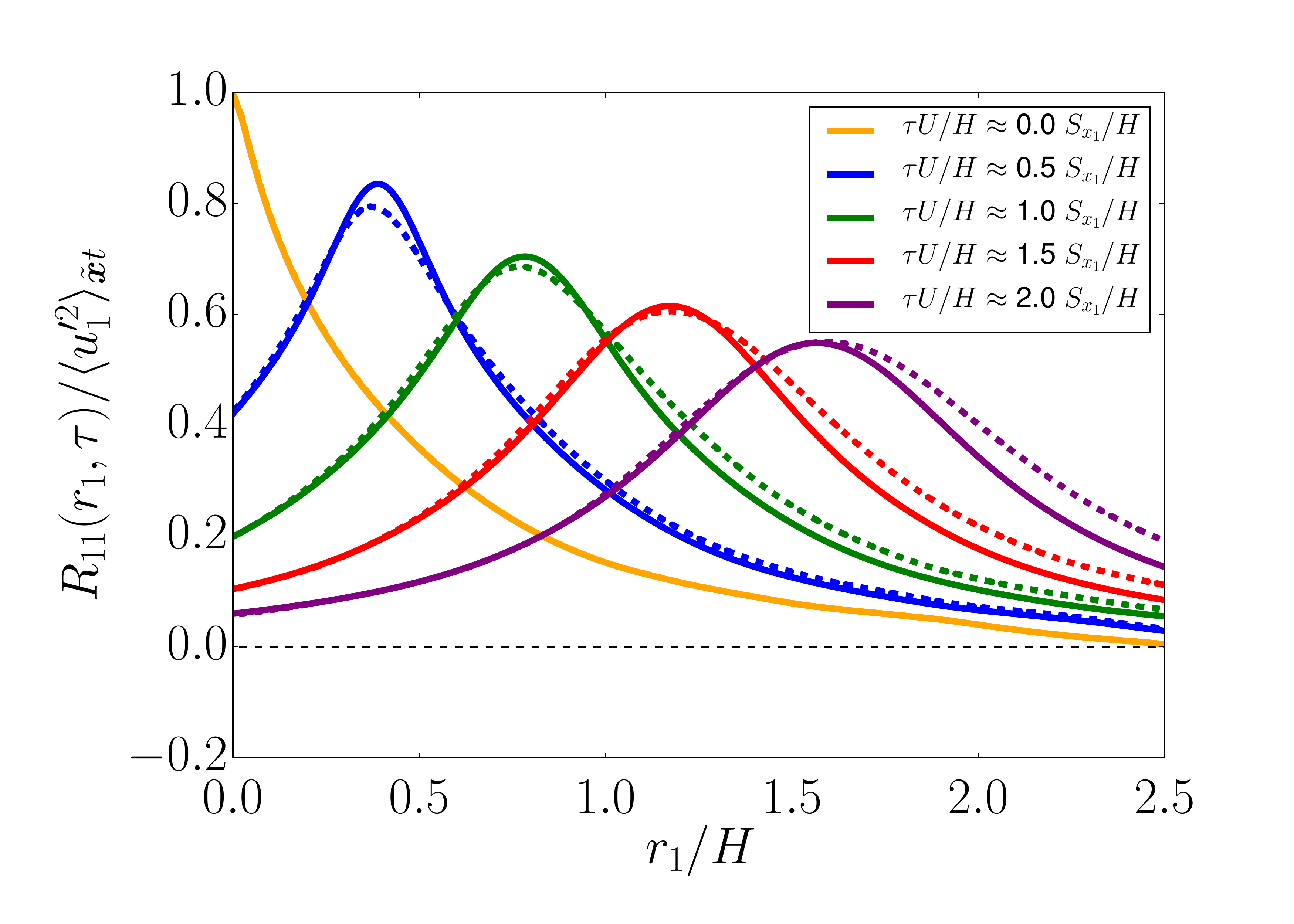}
    \includegraphics[trim = 0.5cm 0.5cm 2.5cm 1.8cm, clip,  width=0.48\textwidth]{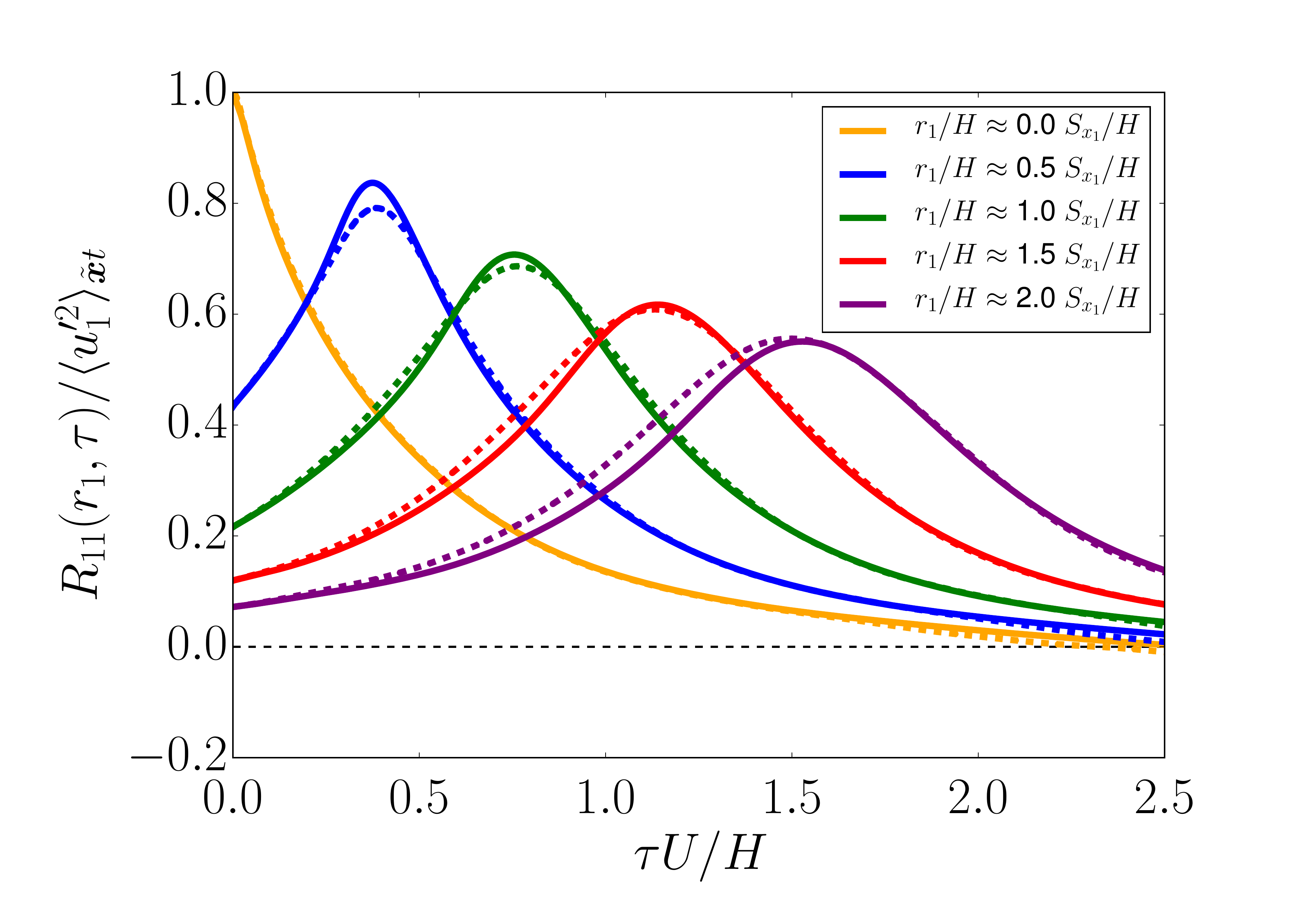}
   \caption{
   Comparison of model to LES data for normalized space-time correlations between the turbine columns. It turns out that with the subsequently described set of parameters excellent agreement can be obtained. Dashed lines correspond to the LES data, solid lines to the normalized model space-time correlation $R_{11}(r_1,\tau)/\langle u_1'^2 \rangle_{\tilde{\bs x} t}$ where the values for $ U  = 9.0\, u^*$ and $\langle v_2^2 \rangle =  0.5\, u^{*2}$ are adjusted for a better accordance with the LES space-time correlation. As above, $R_{11}(r_1)/\langle u_1'^2 \rangle_{\tilde{\bs x} t}$ is evaluated from LES for the lines between the turbines, and the other parameters $\langle v_1^2 \rangle $, $T_1 $, $T_2$, and $L_2 $ are chosen according to case 2 in table \ref{tab:parameters}. Left: normalized $R_{11}(r_1, \tau)$ at fixed instants in time as a function of $r_1$. Right: normalized $R_{11}(r_1, \tau)$ at fixed spatial increments. $S_x=0.785\, H$ is the turbine spacing.}
    \label{fig:figure19_20}
  \end{figure}

To test whether these quantitative discrepancies can be in principle remedied, we evaluate the model for between the turbines with a different set of parameters, only for illustrative purposes. It turns out that by slightly adjusting the mean velocity and the second moment of the spanwise random sweeping, excellent results can be obtained for the lines between the turbines also for large temporal and spatial increments. Figure \ref{fig:figure19_20} shows a comparison to LES for the model space-time correlation with a mean velocity adjusted by a factor $1.14$ to $U=9.0 \, u^*$ and a spanwise variance adjusted by a factor $0.34$ to $ \langle v_2^2 \rangle=0.5\, u^{*2} $. 
In contrast to the streamwise random sweeping, the spanwise random sweeping has influence only on the amplitude but not the width of the model curve. Therefore, we may conclude that the functional form of the proposed model is very satisfactory while the main discrepancies with data are due to deficiencies in prescribing the model parameters. 

\subsection{Discussion}
\label{subsec:discussion}
The comparison of the model \eqref{eq:spacetimecorr} to LES data shows good qualitative agreement in the sense that it captures the main characteristics of spatio temporal velocity correlations in a wind farm, including the spatial and temporal decorrelation and the characteristic tilt in the contour plots generated by an advection velocity. The decorrelation on the lines of turbines is higher than between which is visible in the comparison of the extension of the highly correlated (red) area in the contour plots \ref{fig:figure11_12} and \ref{fig:figure15_16}. The same observation follows from the cuts through the contour plots which show that between the turbine columns, velocity correlations decay slower in space and time, cf.~figure \ref{fig:figure13_14} and \ref{fig:figure17_18}. This is qualitatively reflected by the model.

Considering the fact that all free parameters are obtained from the LES (table \ref{tab:parameters}), the model performs overall well also in terms of quantitative results.
The mean velocity over the respective subsets of lines along the turbines is considerably smaller than between. Still, the choice of the overall mean velocity in our model yields reasonable results on the turbine lines when compared to the LES data. 
One possible explanation could be that velocity fluctuations are not primarily advected with the mean velocity at hub height, but might be entrained from slightly larger heights with a correspondingly higher mean velocity, or influenced by shear effects as suggested in \cite{HeZhang2006}. 

A similar effect became visible in the case between the turbines which in comparison to the LES fits much better for an increased value for the mean velocity. Increasing the value for the mean velocity with respect to the inherent physics of the problem, and at the same time adjusting (by a significant amount) the second moment of the spanwise random sweeping, gives almost perfect overlaps of the curves for between the turbines, cf.~figure \ref{fig:figure19_20}.
Figure \ref{fig:figure17_18} shows that the width of the model space-time correlation between the turbines fits well, however the amplitude is decreasing too fast.
A decreased $\sigma_1$ increases the amplitude and decreases the width of the distribution, a decreased $\sigma_2$ only increases the amplitude and has no influence on the width. This observation suggests that the values chosen for $T_1$ and $\langle v_1^2 \rangle$ are reasonable, whereas a decreased spanwise random sweeping fluctuation $\langle v_2^2 \rangle$ serves to increase the amplitude, as done for figure \ref{fig:figure19_20}. 
In fact, a similar effect would be obtained by decreasing the correlation time of the random sweeping velocity $T_2$. 

Smaller $T_1$ and $T_2$ result in smaller $\sigma_1$ and $\sigma_2$, respectively shorter correlated random sweeping velocities $v_1$ and $v_2$ in equation \eqref{eq:exp_decay_rsvel}. Furthermore, smaller $\sigma_1$ and $\sigma_2$ increase the correlation of $u'_1$ in the model \eqref{eq:spacetimecorr} due to their influence on the amplitude and the width of the decorrelation kernel.
Effectively, this observation gives rise to further discussion with respect to the choice of values for $T_1$ and $T_2$ in the model.
On the one hand, the integral time scales obtained from the LES on the turbine lines are smaller than between the turbine lines, cf.~table \ref{tab:parameters}. This is reasonable considering an expected higher temporal decorrelation of velocity fluctuations on the turbine lines. 
On the other hand, that means that $\sigma_1$ and $\sigma_2$ on the turbine lines are smaller than between the turbine lines. Consequently, with the current choice of values for $T_1$ and $T_2$, the influence of the temporal decorrelation kernel in the model space-time correlation is smaller on the turbine lines than between. However, we leave the discussion of obtaining alternative values for $T_1$ and $T_2$ in the model open for future work. 

To sum up, as already mentioned in the introduction, the temporal decorrelation of the random sweeping velocity has a major influence on the results and we conclude that a temporal decorrelating random sweeping, as incorporated in our model, is necessary to capture the decorrelation more accurately.

\section{Summary and Outlook}
\label{sec:Outlook}

In this paper, we have presented a model for velocity space-time correlations in turbulent flow along with its application to streamwise velocity fluctuations in fully developed wind turbine array boundary layers. The model is based on the classical Kraichnan-Tennekes random sweeping hypothesis with additional mean flow, and takes into account finite correlation times of the large-scale random sweeping velocity. 

Under the given assumptions, the streamwise space-time correlation is obtained as a convolution of the space correlation and a decorrelation kernel, consistent with the classical Kovaznay-Corrsin conjecture. The properties of the analytical decorrelation kernel are fixed by the modeling assumptions, such that the spatial correlation function along with simple statistical characteristics of the velocity field (mean fluctuations, integral time and length scales) suffice to make a prediction for the full space-time correlation.

We then presented detailed comparisons of the model to LES results of a periodic wind farm flow. Taking space-time correlations along the wind turbine columns and between the wind turbine columns as two typical benchmark cases, we find good qualitative agreement given the simplicity of the modeling approach. In this context, it appears worth mentioning that the needed model parameters have been obtained directly from the LES data, such that no free fitting parameters are contained in the formulation. Therefore, the model can be easily adapted to different simulation scenarios and field measurements, which presents one direction of future research.

Regarding limitations of the model, quantitative differences between the model and the LES space-time correlations were observed, especially for the correlations between the turbine columns. These shortcomings are presumably rooted in some oversimplied modeling assumptions as well as the small number of parameters. However, we demonstrated excellent agreement by adjusting the mean velocity and the variance of the spanwise random sweeping. This suggests the potential of the model to capture the space-time correlations quantitatively. One interesting question for future research is therefore how additional physical effects such as vertical energy entrainment or shear affect  the model and its parameters. While vertical energy entrainment may help to justify a higher mean convection velocity, effects like shear can give rise to increased mean velocity fluctuations \cite{HeZhang2006, ZhaoHe2009}. Aiming towards a fully analytical model, it will be interesting to also obtain the spatial correlation function, which is currently obtained from the LES data, from additional theoretical arguments. Such an approach, for example, has been taken in the derivation of the wavenumber-frequency spectrum model with a model wavenumber spectrum \cite{WilczekStevensMeneveau2015}.

Considering power output fluctuations as a next step, both LES results such as \cite{StevensMeneveauRenSus2014} as well as wind tunnel measurements of micro wind farms \cite{BossuytHowlandMeneveauMeyersTorque2016, BossuytMeneveauMeyers2017} may serve as important reference points for future tests of our model.
  
\acks Laura J.~Lukassen and Michael Wilczek are supported by the Max Planck Society.
Richard J.A.M.~Stevens is funded in part by the Shell-NWO/FOM-initiative Computational sciences for energy research of Shell and Chemical Sciences, Earth and Live Sciences, Physical Sciences, FOM and STW. This work was carried out on the national e-infrastructure of SURFsara, a subsidiary of SURF cooperation, the collaborative ICT organization for Dutch education and research. Charles Meneveau was supported by NSF (grant OISE-1243482, the WINDINSPIRE project).

\bibliography{space_time_corr_wind_farms_Submission}{}
\bibliographystyle{wileyj}

\end{document}